%% file: main.tex
\documentclass[twocolumn]{article}
\usepackage[a4paper, textwidth=18cm]{geometry}
\usepackage{graphicx}
\usepackage{framed}
\usepackage[hyphens]{url}      
\usepackage[dvipsnames,table]{xcolor}
\usepackage{mathtools}
\usepackage{lscape}
\usepackage{tabularx}
\usepackage{xurl}
\usepackage{amsthm}
\usepackage{amsmath}
\usepackage{amsfonts}
\usepackage{caption}
\usepackage{subcaption}
\usepackage{authblk}

\usepackage[T1]{fontenc} 
\usepackage{multirow}
\usepackage{tikz}
\usepackage{xpatch}
\usepackage{enumitem}
\usepackage{array}
\usepackage{makecell}
\usepackage{newfloat}
\usetikzlibrary{shapes}
\usepackage{bold-extra}
\usepackage[linesnumbered,vlined]{algorithm2e}
\usepackage{setspace}
\AtBeginDocument{\xpatchcmd\proof{$\blacksquare$}{\qedsymbol}{}{}}
\renewcommand{\qedsymbol}{}

\definecolor{lightviolet}{RGB}{230,200,255}

\SetKw{Continue}{continue}
\SetKw{Break}{break}
\SetKw{True}{true}
\SetKwRepeat{Do}{do}{while}
\SetKwInOut{Input}{Input}
\SetKwInOut{Output}{Output}
\SetKwProg{Fn}{Function}{}{}

\RestyleAlgo{ruled}

\DeclareCaptionType{Example}

\theoremstyle{definition}

\makeatletter
\newcommand\email[2][]%
{\newaffiltrue\let\AB@blk@and\AB@pand
	\if\relax#1\relax\def\AB@note{\AB@thenote}\else\def\AB@note{\relax}%
	\setcounter{Maxaffil}{0}\fi
	\begingroup
	\let\protect\@unexpandable@protect
	\def\thanks{\protect\thanks}\def\footnote{\protect\footnote}%
	\@temptokena=\expandafter{\AB@authors}%
	{\def\\{\protect\\\protect\Affilfont}\xdef\AB@temp{#2}}%
	\xdef\AB@authors{\the\@temptokena\AB@las\AB@au@str
		\protect\\[\affilsep]\protect\Affilfont\AB@temp}%
	\gdef\AB@las{}\gdef\AB@au@str{}%
	{\def\\{, \ignorespaces}\xdef\AB@temp{#2}}%
	\@temptokena=\expandafter{\AB@affillist}%
	\xdef\AB@affillist{\the\@temptokena \AB@affilsep
		\AB@affilnote{}\protect\Affilfont\AB@temp}%
	\endgroup
	\let\AB@affilsep\AB@affilsepx
}
\makeatother
\begin{document}

\input{commands}

\title{
{\utwo: \/Bi-directional Engineering for Hidden Query Extraction}
}

\date{}

\author[1]{Ahana Pradhan}
\author[1]{Jayant Haritsa}

\affil[1]{Indian Institute of Science}
\email{\{\texttt{ahanapradhan,haritsa}\}\texttt{@iisc.ac.in}}

\twocolumn[
  \begin{@twocolumnfalse}
    \maketitle
    \begin{abstract}
      \input{abstract}
    \end{abstract}
  \end{@twocolumnfalse}
]

\maketitle

\input{intro}

\input{background}

\input{design}

\input{union}

\input{aoa}

\input{llm}

\input{expt}

\input{disc}

\input{related_work}

\input{conclusion}

\sloppy

\onecolumn
\bibliographystyle{plain}
\bibliography{main}

\end{document}

%% file: commands.tex
\newcommand{\texttosql}{Text-to-SQL\xspace}

\newcommand{\apptosql}{HQE\xspace}

\newcommand{\atwos}{Text-assisted App-to-SQL\xspace}
\newcommand{\cubes}{{\tt CUBES}\xspace}
\renewcommand{\cellalign}{vh}
\renewcommand\cellgape{\Gape[4pt]}
\definecolor{deepLimeGreen}{rgb}{0.2, 0.6, 0.2}       
\definecolor{lightLimeGreen}{rgb}{0.9, 1.0, 0.9}      

\newcommand{\paper}[1]{#1}

\newcommand{\focus}[1]{\noindent {#1}}
\newcommand{\parttab}[1]{{\tt AuxTables}({#1})}
\newcommand{\fromtab}[1]{{\tt FROM}({#1})}
\newcommand{\wheretab}[1]{{\tt WHERE}({#1})}
\newcommand{\comtab}[1]{{\tt COMMON}({#1})}
\newcommand{\from}{{\sc From}\xspace}
\newcommand{\select}{{\sc Select}\xspace}
\newcommand{\where}{{\sc Where}\xspace}
\newcommand{\unionall}{{\sc Union All}\xspace}
\newcommand{\union}{{\sc Union}\xspace}
\newcommand{\orderby}{{\sc Order By}\xspace}
\newcommand{\groupby}{{\sc Group By}}\xspace
\newcommand{\crossjoin}{{\sc CROSS JOIN}\xspace}
\newcommand{\asc}{{\sc Asc}\xspace}
\newcommand{\desc}{{\sc Desc}\xspace}
\newcommand{\sqland}{{\sc AND}\xspace}
\newcommand{\sqlor}{{\sc OR}\xspace}
\newcommand{\sqlin}{{\sc IN}\xspace}
\newcommand{\qoj}{$Q_{OJ}$\xspace}
\newcommand{\sigmaap}{${\sigma}_{AP}$}
\newcommand{\sigmalp}{${\sigma}_{ALP}$}
\newcommand{\alp}{{ALP\xspace}}
\newcommand{\coretables}{{\tt CoreTables\xspace}}

\newcommand{\sidetables}{{\tt SideTables\xspace}}

\newcommand{\tpch}{{TPCH\xspace}}
\newcommand{\xdata}{{XData\xspace}}
\newcommand{\etpch}{{E-TPCH\xspace}}
\newcommand{\stack}{{STACK\xspace}}
\newcommand{\aux}{{\tt Aux\xspace}}
\newcommand{\tq}{{$TX_Q$}\xspace}
\newcommand{\dnep}{$D_{nep}$\xspace}
\newcommand{\appl}{$\mathcal{A}$\xspace}
\newcommand{\utwobasic}{{\tt XPOSE}$_{basic}$\xspace}
\newcommand{\QH}{$Q_{\mathcal{H}}$\xspace}
\newcommand{\tH}{$T_{\mathcal{H}}$\xspace}
\newcommand{\QE}{$Q_{\mathcal{E}}$\xspace}
\newcommand{\QS}{$Q_{\mathcal{S}}$\xspace}
\newcommand{\RH}{$\mathcal{R}_{\mathcal{H}}$\xspace}
\newcommand{\RE}{$\mathcal{R}_{\mathcal{E}}$\xspace}
\newcommand{\RS}{$\mathcal{R}_{\mathcal{S}}$\xspace}
\newcommand{\fullresult}{\RH\xspace}
\newcommand{\todo}{{\color{red} To-do}}
\newcommand{\D}{$D_I$\xspace}
\newcommand{\Dcap}{$\widehat{D_I}$}
\newcommand{\done}{$D^1$\xspace}
\newcommand{\dmin}{$D_{min}$\xspace}
\newcommand{\parttabc}[1]{$\overline{\mathtt{AuxTables}}$({#1})}
\newcommand{\uone}{{\sc Unmasque}\xspace}
\newcommand{\utwo}{{\sc Xpose}\xspace}
\newcommand{\op}{$op$}
\newcommand{\filterex}{SVE\xspace}
\newcommand{\filtergrp}{$GS_{ve}$\xspace}
\newcommand{\svalinterval}{$SVI$\xspace}
\newcommand{\mybinom}[2]{\Bigl(\begin{array}
{@{}c@{}}#1\\#2\end{array}\Bigr)}
\newcommand{\cost}[2]{{\em Cost}({#1}) }
\newcommand{\disrel}[1]{{\em Dis}({#1})}
\newcommand{\conrel}[1]{{\em Con}({#1})}
\newcommand{\xmark}{\ding{55}}%
\newcommand{\nullo}{NULL\xspace}
\newcommand{\pg}{{PostgreSQL}\xspace}

\newcommand*{\balancecolsandclearpage}{%
  \close@column@grid
  \cleardoublepage
  \twocolumngrid
}
\newcommand{\STAB}[1]{\begin{tabular}{@{}c@{}}#1\end{tabular}}

%% file: abstract.tex
Query reverse engineering (QRE) aims to synthesize a SQL query to connect a given database and result instance. A recent variation of QRE is where an additional input, an {\em opaque executable} containing a ground-truth query, is provided, and the goal is to non-invasively extract this specific query through only input-output examples.  This variant, called Hidden Query Extraction (HQE), has a spectrum of industrial use-cases including query recovery, database security, and vendor migration.

The reverse engineering (RE) tools developed for HQE, which are based on database mutation and generation techniques, can only extract flat queries with key-based equi-joins and conjunctive arithmetic filter predicates, making them limited wrt both query structure and query operators. 
In this paper, we present {\utwo}, a HQE solution that elevates the extraction scope to realistic complex queries, such as those found in the {\tpch} benchmark.

A two-pronged approach is taken: (1) The existing RE scope is substantially extended to incorporate \emph{union} connectors, \emph{algebraic} filter predicates, and \emph{disjunctions} for both values and predicates. (2) The predictive power of LLMs is leveraged to convert business descriptions of the opaque application into extraction guidance, representing ``forward engineering'' (FE). The FE module recognizes common
query constructs, such as \emph{nesting} of sub-queries, \emph{outer joins}, and \emph{scalar functions}. In essence, FE establishes the broad query contours, while RE fleshes out the fine-grained details.

We have evaluated \utwo on (a) {\etpch}, a query suite comprising the complete {\tpch} benchmark extended with queries featuring unions, diverse join types, and sub-queries; and (b) the real-world {\stack} benchmark. The experimental results demonstrate that its bi-directional engineering approach accurately extracts these complex queries, representing a significant step forward with regard to HQE coverage.

%% file: intro.tex
\section{Introduction}
Query Reverse Engineering (QRE) has been a subject of considerable recent interest in the database community. Here, the standard database equation, namely \emph{Q(\D) = {$\mathcal{R}$}}, where Q is a declarative query, \D is a database instance, and {$\mathcal{R}$} is the result of executing Q on {\D} -- is inverted. Specifically, the inputs are now \D and $\mathcal{R}$, and the goal is to identify a candidate SQL query $Q_c$ that satisfies the equation. QRE has a variety of industrial applications, including recreating lost application code and assisting SQL amateurs to formulate complex queries. To meet these objectives, a host of sophisticated software tools (e.g.~TALOS~\cite{talos}, Scythe \cite{scyth}, FastQRE~\cite{fastqre}, REGAL~\cite{regal_plus, regal},  SQUARES~\cite{squares}, PATSQL~\cite{patsql}, SICKLE~\cite{qbe_application}, CUBES ~\cite{cubes}) have been developed over the past decade.

\subsection*{Hidden Query Extraction}
A variant of QRE, called Hidden Query Extraction (HQE), was introduced in Sigmod 2021~\cite{unmasque_sigmod}. Here, in addition to the generic QRE inputs, \D and $\mathcal{R}$,  a ground-truth query is available, but in a hidden or inaccessible form. For instance, the query may have been encrypted or obfuscated. The goal now is to non-invasively extract the hidden query through ``active learning'', that is, by observing a series of \emph{input-output examples} produced by this executable on a variety of curated databases.~\footnote{This problem is akin to the classical ``Chosen Plain-text Attack'' in cryptography~\cite{cpa}.}

HQE has several use-cases, including application logic recovery, database security, vendor migration, and query rewriting. For instance, with legacy industrial applications, the original source code is often lost with passage of time~\cite{shf, StackExchangeLosingSourceCode} 
(particularly with organizational mergers or outsourcing of projects). 
However, to understand the application output, we may need to establish the logic connecting the database input to the observed result. Moreover, we may wish to extend or modify the existing application query, and create a new version.

Formally, the HQE problem is defined as: \emph{Given a black-box application {\appl} 
containing a hidden SQL query \QH, and a sample database instance \D on which {\appl} 
produces a populated result {\fullresult} (i.e. \QH (\D) = {\fullresult}),  determine the precise \QH contained in {\appl}.
}  

At first glance, 
HQE may appear more tractable than general QRE since the executable could be leveraged to prune the candidate search space. But, conversely, the correctness requirements are \emph{much stricter} since a specific ground-truth query, and not a candidate solution, is expected as the output. So, while the problems are related, their solution frameworks are quite different.

\subsubsection*{{{\uone}} Extractor}
We took a first step towards addressing the HQE problem in \cite{unmasque_sigmod}, by introducing the {\bf UNMASQUE} tool, which is capable of (non-invasively) extracting flat {\bf SPJGAOL} {\sc(Select, Project, Join, Group By, Agg, Order By, Limit)} queries. The extraction operates in the linear pipeline shown in Figure~\ref{fig:unmasque}.
Here, a clause-by-clause extraction of the hidden query is implemented, starting with \from and ending with {\sc Limit}. This is achieved by analyzing the results of targeted \QH executions on databases derived from \D through {\em database mutation} and {\em database generation} techniques. To reduce extraction overheads, the pipeline begins by minimizing the sample input \D to a handful of rows.

\input{u1_pipeline}

\subsection*{Bi-directional Engineering}
A natural question at this point is whether the extraction scope of \uone could be extended to ``industrial-strength'' queries. For instance, as a tangible milestone, is it feasible to extract the {\tpch} benchmark queries, most of which go well beyond flat SPJGAOL? Our analysis shows that, by moving from a linear extraction pipeline to a \emph{reentrant} pipeline with looping across modules, the RE-based techniques can be extended to include complex operators. In particular, Unions and Algebraic Predicates (\emph{column op column}) are described later in this paper. But, even so, we also found them to be fundamentally incapable of extracting common query constructs, such as nestings (in {\sc From} and {\sc Where} clauses) and scalar functions (e.g. {\tt substring()}).

The above limitations appear to preclude the extraction of realistic queries. However, this negative outcome can be remedied if we assume that with the executable, a high-level \emph{textual description}, \tq, of the business logic embedded in the query is also available. Such information is usually present in application design specifications, and therefore likely to be accessible. In fact, industries subject to compliance regulations (e.g. banking, insurance, healthcare) are \emph{mandated} to possess documented business logic for audit purposes.

We can now bring in the remarkable predictive power of LLMs to leverage \tq into extraction guidance -- that is, ``forward engineering'' (FE), akin to Text-to-SQL tools~\cite{text_to_sql_paperswithcode}, specifically for features such as query nesting, outer joins, and scalar functions. And what we show in this paper is that a judicious synergy of forward and reverse engineering is capable of precisely extracting the \emph{entire {\tpch} benchmark} and more. Moreover, this coverage could \emph{not} have been achieved by either of them in isolation. In a nutshell, FE establishes the broad contours of the query, while RE fleshes out the fine-grained details in the SQL clauses.

\subsection*{The {\utwo} System}
\begin{figure}[!h]
    \centering
    \includegraphics[width=0.9\linewidth]{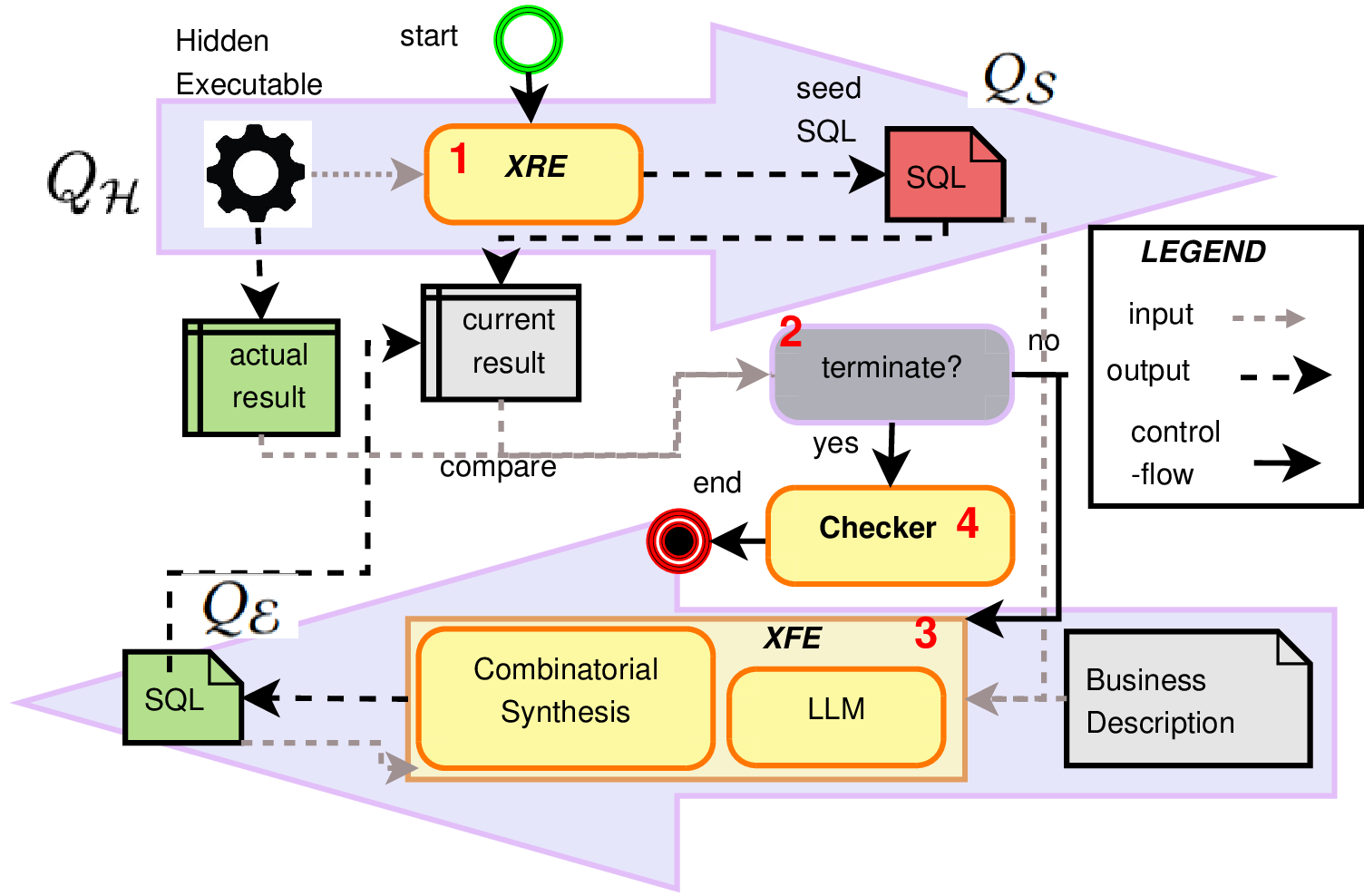}
    \caption{Architecture of {\utwo}}
    \label{fig:xpose_problem}
\end{figure}

Our new bidirectional approach is implemented in a tool called {\utwo}, depicted in Figure~\ref{fig:xpose_problem}.  The inputs are the textual description \tq, the opaque executable \appl containing \QH, and the sample database \D.
The extraction process starts with the {\bf XRE} {\em (\utwo-RE)} module, which uses \appl and a variety of databases derived from \D to reverse-engineer a {\em seed query}, \QS. 
Then \QS is sent to the {\bf XFE} {\em (\utwo-FE)} module, which contrasts it against \tq, and iteratively refines it, using feedback prompting techniques, into \QE, a semantically equivalent version of {\QH} through query synthesis. The refinement iterates until one of the following occurs: (1) the synthesized query matches \QH wrt the execution results on \D; (2) no new formulations are synthesized; or (3) a threshold number of iterations is exceeded. If the LLM stops due to the latter two cases, which represent failure, we move on to a {\em combinatorial synthesizer} that searches through all valid combinations of tables and groupings within the nesting structure constructed by the LLM. 
Finally, although fundamentally limited due to non-availability of the hidden query, we try various tests to check the accuracy of the extraction: First, for queries within its scope, the {\xdata}~\cite{xdata} tool is used to generate test databases that ``kill mutants'' -- that is, produce different results if there are discrepancies between the extracted query \QE and the black-box \QH. 
Second, the result-equality of \QE and \QH is evaluated on multiple randomized databases to probabilistically minimize the likelihood of false positives. 

\input{case}

\subsection*{Illustrative Extraction}
To illustrate the above process, consider TPCH {\bf Q20} shown in Figure~\ref{fig:usecases}(a) with encryption. The business description~\cite{tpch} of this so-called ``Potential Part Promotion Query'', is in Figure~\ref{fig:usecases}(b).

The XRE output of \utwo with the Q20 executable is shown in Figure~\ref{fig:usecases}(c). We see here that it correctly extracts the five base tables ({\sc Lineitem}, {\sc Nation}, {\sc Part}, {\sc Partsupp}, {\sc Supplier}), equi-joins among these tables, filter predicates ({\tt l\_shipdate}, {\tt p\_name}, {\tt n\_name}), projection columns, and result ordering. On the downside, however, there are several errors (highlighted in red) -- for instance, filters on {\tt l\_quantity} and {\tt ps\_availqty} are \emph{spuriously} reported due to the nested subquery in the \where clause, which goes beyond the flat structures covered by XRE.  Similarly, the joins within sub-queries are incorrectly characterized as global joins. Lastly, semi-joins, constructed using subqueries, are extracted as equi-joins.

The XFE component is now prompted with (i) the business description \tq,
(ii) the XRE output as a seed query \QS, and (iii) a set of canonical correction instructions listed in Figure~\ref{fig:usecases}(d). These instructions are generic guidelines, based on SQL experience and XRE limitations, to help the LLM synthesize queries that are aligned with \QS and \tq. As a simple instance, the LLM's schematic scope is explicitly restricted to the tables in \QS since they are known to have been identified correctly. Ensuring compliance with this suite of instructions results in correction of all previously noted errors, as shown in the final extracted query, \QE, listed in Figure~\ref{fig:usecases}(e) (the respective roles of XRE and XFE are highlighted in blue and purple colors). 
Finally, the checker module is invoked to test whether, modulo syntactic differences, \QE is functionally equivalent to \QH -- for Q20, both the {\xdata} tool and the result-equivalence tests do not find any discrepancy.

\subsection*{Contributions} 
In this paper, our contributions are with regard to both forward and reverse engineering:

\begin{itemize}[wide, labelwidth=!, labelindent=0pt]
\item  Constructing RE algorithms to extract unions, algebraic predicates, and predicate disjunctions, by generalizing the linear pipeline of {\uone} to a reentrant directed graph.

\item FE using an LLM on a seed query augmented with automated corrective feedback prompts. The prompts are designed to rectify the errors (of both commission and omission) introduced by RE. 

\item 
A novel amalgamation of FE and RE to extract complex SQLs, implemented in \utwo.

\item Evaluation of {\utwo} on (a) {\etpch}, a query suite featuring the complete {\tpch} benchmark extended with queries capturing diverse join types and unions of sub-queries; and (b) {\stack} benchmark queries featuring several joins and disjunction predicates.

\end{itemize}

\subsection*{Organization} The mutation framework of \uone is reviewed in Section~\ref{sect:background}.  The high-level design of \utwo is described in Section~\ref{sect:design}, followed by the building blocks of XRE and XFE. Detailed XRE extraction strategies for union and algebraic predicates are presented in Sections~\ref{sect:union} and \ref{sect:aoa}, respectively. The use of XFE to synthesize from the seed query is detailed in Section~\ref{sect:llm}. The experimental framework and results are presented in
Section~\ref{sect:expt}. Our future research avenues are discussed in Section~\ref{sect:discuss}, and related work is reviewed in Section~\ref{sect:relwork}.
Finally, our conclusions are summarized in Section~\ref{sect:conclusion}.

%% file: u1_pipeline.tex
\begin{figure}[!h]
  \centering
  \includegraphics[width=0.85\columnwidth]{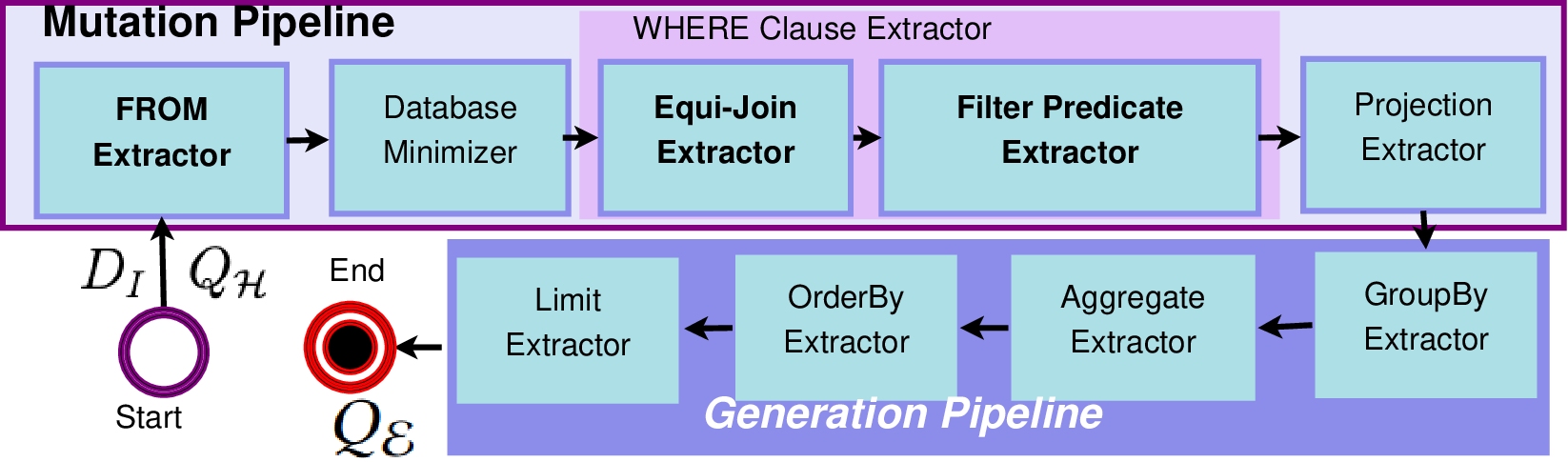}
  \caption{Linear Extraction Pipeline of {\uone} \cite{unmasque_sigmod, kapil_demo}}
  \label{fig:unmasque}
\end{figure}

%% file: case.tex
\begin{figure*}[h!bt]
{\footnotesize
\begin{tabular}{p{11cm}p{6cm}}
(a) {\small\bf Hidden Query: \QH \hspace*{0.1in} [Encrypted {\tpch} Q20]}&(b) {\small\bf Business Description of Q20: \tq}\\
{\bf CREATE PROCEDURE hQ With Encryption BEGIN} {\color{black}select
        s\_name,
        s\_address
from
        supplier,
        nation
where
        s\_suppkey in (
                select
                        ps\_suppkey
                from
                        partsupp
                where
                        ps\_partkey in} (
                                {\color{black}select
                                        p\_partkey
                                from
                                        part
                                where
                                        p\_name like '\%ivory\%'}
                        )
                        {\color{black}and ps\_availqty >} (
                                select
                                        0.5 * sum(l\_quantity)
                                from
                                        lineitem
                                where
                                        l\_partkey = ps\_partkey
                                        and l\_suppkey = ps\_suppkey
                                        and l\_shipdate >= date '1995-01-01'
                                        and l\_shipdate < date '1995-01-01' + interval '1' year
                        )
        )
        {\color{black}and s\_nationkey = n\_nationkey
        and n\_name = 'FRANCE'
order by
        s\_name} {\bf END};
&
\textcolor{black}{The query identifies suppliers who have an excess of a given part available}; {an excess is
defined to be more than 50\% of the parts like the given part that the supplier shipped in the year 1995 for FRANCE}. {\color{black}Only parts whose names share the word `ivory' are considered.}\\\\

\end{tabular}
\begin{tabular}{p{7.5cm}p{9.5cm}}
 (c) {\small\bf Output of XRE (Seed Query): \QS}&{\small \bf (d) Feedback Prompts to LLM} \\
 \textcolor{black}{Select s\_name, s\_address 
 From} lineitem, {\color{black}nation}, part, {\color{black}partsupp, supplier} 
 Where l\_partkey = p\_partkey
 and {\color{black}p\_partkey} = {\color{black}ps\_partkey}
 and l\_suppkey = ps\_suppkey
 and {\color{black}ps\_suppkey = s\_suppkey
 and n\_nationkey = s\_nationkey
 and n\_name = 'FRANCE'}
 and {\color{red}l\_quantity <= 9687.99}
 and l\_shipdate between '1995-01-01' and '1995-12-31'
 and {\color{black}p\_name LIKE '\%ivory\%'}
 and {\color{red}ps\_availqty >= 12} 
 {\color{black}Order By s\_name asc};         &

{\footnotesize
(1) Only use the tables in the seed query.  
(2) Validate all the filter predicates in the seed query against the text.
(3) Use the valid filter predicates present in the seed query. 
(4) For the attributes in the invalid filter predicates, validate their use from the business description text. Then formulate predicates with them.
(5) Strictly follow the projection order and projection dependencies used in the seed query.
(6) A semi-join, implying at least one match, may be incorrectly present as an equi-join in the seed query.}
\\\\
\multicolumn{2}{p{17.5cm}}{{\small{\bf (e) Final Output of \utwo (\textcolor{blue}{XRE} + \textcolor{purple}{XFE}): \QE}}} \\
\multicolumn{2}{p{17.5cm}}{
\footnotesize
\textcolor{blue}{Select s\_name, s\_address
From supplier
Where s\_suppkey} \textcolor{purple}{IN (
    Select} \textcolor{blue}{ps\_suppkey
    From partsupp
    Where ps\_availqty} \textcolor{purple}{> 0.5 * (
        Select Sum(}\textcolor{blue}{l\_quantity}\textcolor{purple}{)}
        \textcolor{blue}{From lineitem
        Where l\_shipdate BETWEEN `1995-01-01' AND `1995-12-31'
        and l\_partkey = ps\_partkey
        and l\_suppkey = ps\_suppkey}
    \textcolor{purple}{)}
    \textcolor{blue}{and ps\_partkey} \textcolor{purple}{IN (
        Select} \textcolor{blue}{p\_partkey
        From part
        Where p\_name LIKE '\%ivory\%'}
    \textcolor{purple}{)
)}
\textcolor{blue}{and s\_nationkey = (
    Select n\_nationkey
    From nation
    Where n\_name = 'FRANCE'
)}
\textcolor{blue}{Order By s\_name ASC};}\\

\end{tabular}
}
    \caption{Extracting {\tpch} Q20 using Bidirectional Engineering}
    \label{fig:usecases}

\end{figure*}

%% file: background.tex
\section{Mutation Framework of \uone~\cite{unmasque_sigmod}}
\label{sect:background}
We review here a few core concepts of the mutation framework employed in \uone. These concepts are augmented in \utwo, as described in Section~\ref{sect:design}. The notations used here and in the rest of the paper are listed in Table~\ref{tab:notations}.

\input{notation}

To extract the SPJ-elements (of flat SPJGAOL queries), \uone employs
targeted data mutations at different granularities (relation or attribute) to expose clause-specific unique signatures.
The underlying assumptions are: (1) The 
database \D is NULL-free; (2) \QH produces a populated result on \D; 
(3) \QH is a conjunctive query.
The simple query shown in Figure~\ref{fig:uonescope} is used to illustrate the {\uone} concepts.

\begin{figure}[!h]
\begin{subfigure}[b]{\linewidth}
    \noindent{\small
    \begin{tabular}{p{8cm}}
 $q_0$: 
\select
c\_name AS name, c\_phone as phone
\from customer, orders	\where c\_custkey = o\_custkey \sqland c\_acctbal <= 10000\\
    \end{tabular}
    }
    \caption{Simple Hidden Query $q_0$}
    \label{fig:uonescope}
\end{subfigure}

\vspace*{0.1in}

 \begin{subfigure}[b]{\linewidth}       
\smallskip
\begin{center}
{\footnotesize

\begin{tabular}{|c|c|c|c|c|c|}
\multicolumn{6}{c}{\bf Minimized Database \done}\\
\multicolumn{2}{c}{Table: \done.{\sc Orders}}&\multicolumn{1}{c}{}& \multicolumn{3}{c}{Table: \done.{\sc Customer}}\\\cline{1-2}\cline{4-6}
{$o\_custkey$}&$\dots$ &&{$c\_acctbal$}&$c\_custkey$ &$\dots$\\\cline{1-2}\cline{4-6}
 23074 &$\dots$& &774.84 &23074 &$\dots$\\
\cline{1-2}\cline{4-6}
\end{tabular}
}

\vspace*{0.1in}

{\footnotesize
\begin{tabular}{|c|c|}
\multicolumn{2}{c}{{\bf Populated Result Set obtained by \QH on \done:}}\\\hline
{\em name} & {\em phone}\\\hline
Customer\#000023074&18-636-637-7498\\\hline
\end{tabular}

}
\end{center}
\smallskip

\caption{\done and result with $q_0$ on TPCH database}
    \label{fig:si_se}
    \end{subfigure}
    \caption{Database Minimization}
    \vspace*{-0.2in}
\end{figure}

\subsection{From Clause Extractor}\label{sect:from}
The \from clause extractor enumerates {\tH}, the set of tables present in \QH. For this purpose, a simple {\bf Extraction by Error (EbE)} check is sequentially carried out over all tables in the database schema. Specifically, a table is present if renaming it to a dummy name triggers an error from the database engine when parsing {\QH}. The original database schema is restored after each check is completed. 
For $q_0$, EbE produces {\tH} = \{{\sc Customer}, {\sc Orders}\}.

An alternative check, {\bf Extraction by Voiding (EbV)}, determines a table is present if its ``void'' version (i.e. table having no data) causes \RH to be empty. This check was intended to be used in \uone only for certain special cases where EbE did not apply, but is routinely invoked in \utwo for extracting Unions of sub-queries (Section~\ref{sect:union}).

\subsection{Database Minimizer}\label{sect:db_min}
To avoid long extraction times due to repeated executions of \QH on a large {\D},
the notion of a minimized database \dmin was introduced in \uone.
Specifically, a database is said to be minimal if removing a row from any table leads to an empty result. That is, it is the minimal database that provides a populated output. 
Interestingly, it was proved in \cite{unmasque_sigmod} that under the aforementioned assumptions, there always exists a \dmin wherein each table in {\tH} contains only a \emph{single row}. Such a database is referred to as $\mathbf D^1$. It is efficiently identified using a recursive halving process~\cite{unmasque_sigmod}.

A sample \done for $q_0$, computed from the {\tpch} database, is shown in Figure~\ref{fig:si_se}, along with the corresponding result set.

\subsection{Arithmetic Filter Predicate Extractor}\label{sect:where}

Arithmetic filter predicates of the form \emph{column op value}, where $op \in \{=, <, \leq, \geq, >\}$, are extracted with the following process: For each attribute of the tables in {\tH},  a binary search-based mutation on \done is carried out over its domain [$i_{min}$, $i_{max}$]. 
In Figure~\ref{fig:si_se}, \done.$c\_acctbal$ $= 774.84$, and our goal is to find whether predicate $l \leq$ $c\_acctbal$ $\leq r$ exists on $c\_acctbal$.
\done is first mutated with $c\_acctbal$ $= i_{min}$, and $q_0$ is executed. A populated result implies $i_{min} \leq$ $c\_acctbal$ $\leq 774.84$ satisfies the predicate, i.e. $l = i_{min}$. 
Next, \done is mutated with $c\_acctbal$ $= i_{max}$, which causes $q_0$ to produce an empty result. It implies $774.84 \leq$ $c\_acctbal$ $<$ $r$ $<$ $i_{max}$, for some $r$. So, a binary search is mounted on
the interval [774.84, $i_{max}$] to identify $r$, which turns out to be 10000.00.

\subsection{Satisfying Values}
\label{sect:sv}
The term ``Satisfying values'', or \emph{s-values}, was introduced to refer, for each attribute, the range of its values that satisfy \QH and contribute towards producing a populated result.
The predicate filter values determine the s-values for the corresponding attribute. For instance, the s-values for $c\_acctbal$ is the closed interval [$i_{min}$, 10000.00].

%% file: notation.tex
\begin{table}[!h]
    \centering
    {\footnotesize
    \begin{tabular}{|c|l|}
        \hline
        {\em Symbol} & {\em Meaning} \\\hline\hline
        {\appl} & Opaque Executable \\\hline
        {\tq} & Business Description Text \\\hline
        {\D} & Initial Database \\\hline
        {\done} & Single Row Database \\\hline
        {\QH} & Hidden Query \\\hline
        {\fullresult} & Output of \QH \\\hline
        {\QS} & Seed query extracted by XRE \\\hline
        {\RS} & Output of \QS \\\hline
        {\QE} & Final query extracted by \utwo \\\hline
        {\RE} & Output of \QE \\\hline
        {\tH} & Set of tables in \QH \\\hline
        {ALP} & Algebraic Predicates \\\hline
        $i_{min}$ & Min Domain Value \\\hline
        $i_{max}$ & Max Domain Value \\\hline
        {\filterex} & S-value Extractor \\\hline
        {\svalinterval} & S-value intervals \\\hline
        {\svalinterval}$_{c}^{UB}$ & Max s-value of attribute $c$ \\\hline
        {\svalinterval}$_{c}^{LB}$ & Min s-value of attribute $c$ \\\hline
    \end{tabular}
    \caption{Notations}
    \label{tab:notations}
    }
\end{table}

%% file: design.tex
\section{Design Overview of \utwo}\label{sect:design}
In this section, we provide an overview of the \utwo design. We begin with the distribution of work across the XRE and XFE modules, then follow up with how the mutation concepts of \uone (Section~\ref{sect:background}) are adapted in XRE to meet the significantly extended extraction scope requirements of \utwo, and conclude with the synthesis framework of XFE.

\begin{figure}[!h]
    \centering
    \includegraphics[width=0.7\linewidth]{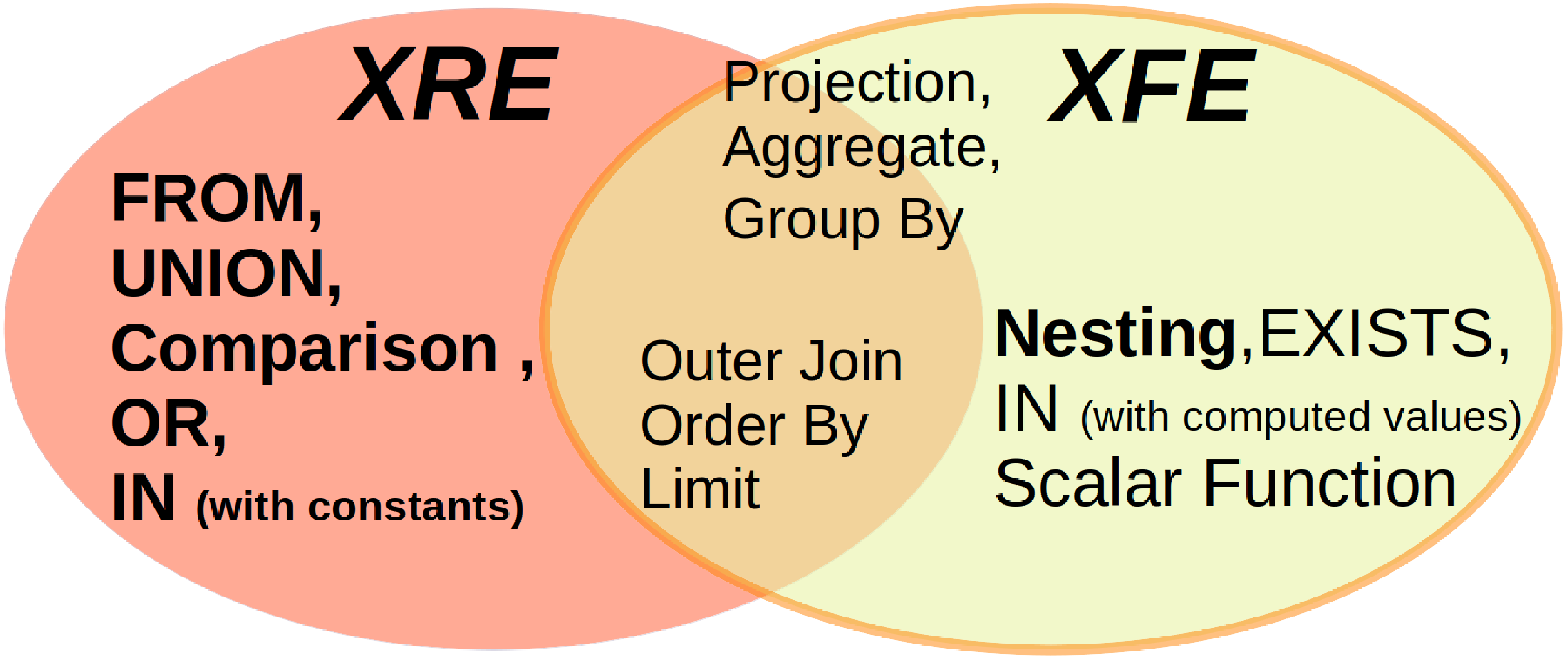}
    \caption{Extraction Scopes of XRE and XFE}
    \label{fig:scope_division}
\end{figure}

\subsection{XRE and XFE: Who does What?}
An interesting design question is the division of extraction labor between XRE and XFE. For instance, since we already know that XRE has some inherent limits, can we go to the other extreme and carry out most or all of the extraction using XFE? The answer is no, we need a \emph{combination} of XRE and XFE that play to their respective strengths, to achieve precise extraction of complex queries. Specifically, XRE is deterministic and provable, whereas XFE is iterative and probabilistic, requiring XRE as an anchor. 

As a case in point, in Figure~\ref{fig:usecases}, if we employ XFE to directly carry out the synthesis from only \tq, several errors are made -- a spurious introduction of the {\sc Region} table in the outer query, creation of two instances of {\sc Lineitem}, applying temporal predicate on {\tt o\_orderdate} instead of {\tt l\_shipdate}, using wrong attribute {\tt l\_quantity} in the aggregate comparison, etc. 
However, armed with the XRE-generated seed query of Figure~\ref{fig:usecases}(c), XFE is able to restrict its formulation to the precise set of tables and attributes, and converge the synthesis to \QH.

Our division of labor between the XRE and XFE components is shown in Figure~\ref{fig:scope_division}. Only XRE can guarantee correct extractions of operators performing core associations among the data items, such as {\union}, {\from}, equality predicates (including {\sc Joins}), and disjunctive ({\sqlor} and {\sqlin} operator with query constant values) predicates. On the other hand, nested structures, membership operators (such as {\sqlin} operator with query computed values), and scalar functions can only be handled by XFE. Whereas operators such as \select, {\sc Agg}, and {\groupby} require a \emph{combined} effort -- for instance, in projection functions, XRE identifies the variables (attributes) and XFE formulates the function (such as polynomial and conditional computations) linking these variables. Finally, 
the comparison operators, {\sc Outer Join}, \orderby and {\sc Limit} clauses, can all be extracted by \emph{either} XRE or XFE. For these operators, the advantage of XRE is provable correctness whereas that of XFE is performance benefits (due to immediate derivation from \tq). In our implementation, we chose to have comparisons handled by XRE (because {\sc Joins} and comparisons are handled uniformly by our algebraic predicate extractor), whereas the remainders are delegated to XFE.

\subsection{Running Example of Hidden Query}\label{sect:running}
To explain the working of \utwo, hereafter we use the hidden query shown, along with its business description, in Figure~\ref{fig:utwoscope}. This query has two sub-queries, $q_1$ and $q_2$, and is designed to highlight the major new extraction extensions provided by {\utwo} -- nested structures (nested select in $q_2$), unions of sub-queries (\unionall of $q_1$ and $q_2$), algebraic predicates  
(${\tt s\_acctbal} \leq {\tt o\_totalprice}$ in $q_2$),  disjunctions ({\sqlin} in $q_2$), and outer joins (in $q_1$).

\begin{figure}[!ht]
\begin{subfigure}[b]{\columnwidth}
\centering

\noindent{\footnotesize
\begin{tabular}{p{8cm}}\hline
{\em Business Description} \tq: List the names and phone numbers of customers and suppliers with low balance. For customers having orders, low balance is determined by a constant threshold whereas for suppliers, the threshold is determined by any order shipped using specified transport modes. Customers having no orders are also included in this list, irrespective of their balance.\\\hline
\medskip
\QH: 
\select * \from ( \\
\begin{tabular}{@{}l l@{}}
\textcolor{blue}{\large$q_1$} &
\fcolorbox{deepLimeGreen}{lightLimeGreen}{%
\begin{minipage}{7.5cm}
 (  \select c\_name AS name, c\_phone AS phone \\
\hspace*{0.5cm}{\from} {\sc Customer} \textcolor{blue}{\sc Left Outer Join} {\sc Orders}\\
\hspace*{1cm}\textcolor{blue}{\sc On} c\_custkey = o\_custkey\\
\hspace*{1cm} \where c\_acctbal <= 10000 OR o\_orderkey {\sc Is} {\sc Null} 
)
\end{minipage}
}\\
\end{tabular}\\
\hspace*{0.5cm} \textcolor{blue}{\textbf{UNION ALL}} \\

\begin{tabular}{@{}l l@{}}
\textcolor{blue}{\large$q_2$} &
\fcolorbox{deepLimeGreen}{lightLimeGreen}{%
\begin{minipage}{7.5cm}
 ( \select s\_name AS name, s\_phone AS phone  \from supplier \\
\hspace*{0.3cm} \where s\_suppkey \textcolor{blue}{\sc In (} \select l\_suppkey \from orders, lineitem \\
\hspace*{0.6cm} \where l\_orderkey = o\_orderkey \sqland \textcolor{blue}{s\_acctbal <= o\_totalprice} \\
\hspace*{0.6cm} \sqland \textcolor{blue}{l\_commitdate = l\_receiptdate}\\  \hspace*{0.6cm}\sqland l\_shipmode \textcolor{blue}{\sc In ('AIR', 'TRUCK')}\textcolor{blue}{)} 
) 
\end{minipage}
}\\
\end{tabular}\\
\hspace*{0.2cm} ) AS people {\groupby} name, phone \\
\end{tabular}
}
\caption{Hidden Query \QH with Business Description}
\label{fig:utwoscope}
\end{subfigure}

\bigskip

 \begin{subfigure}[b]{\columnwidth}
     \centering
     \noindent{\footnotesize
    \begin{tabular}{p{8cm}}
 \QS:   (
\select
c\_name AS name, c\_phone as phone
\\\hspace*{0.5cm} \hspace*{0.5cm}\from customer, orders
	\where c\_custkey = o\_custkey \sqland c\_acctbal <= 10000.00 OR o\_orderkey {\sc Is} {\sc Null} {\groupby} c\_name, c\_phone
)\\
{\unionall} \\ \hspace*{0.5cm}(
\select s\_name AS name, s\_phone as phone\\
	\hspace*{0.5cm} \hspace*{0.5cm}\from lineitem, orders, supplier 
			   \where l\_orderkey = o\_orderkey \sqland s\_suppkey = l\_suppkey
			   \sqland s\_acctbal <= o\_totalprice\\ 
			   \hspace*{0.5cm} \hspace*{0.5cm}\sqland l\_commitdate = l\_receiptdate\\\hspace*{0.5cm} \hspace*{0.5cm}\sqland l\_shipmode {\sqlin} ('AIR', 'TRUCK')\\
                \hspace*{0.5cm}{\groupby} s\_name, s\_phone
	)\\
    \end{tabular}
    }
\caption{Seed Query \QS (output by XRE)}
     \label{fig:xre_case}
 \end{subfigure}

\bigskip

 \begin{subfigure}[b]{\columnwidth}
     \centering
     \noindent{\footnotesize
    \begin{tabular}{p{8cm}}
 \QE:   
\select name, phone 
\from ( 
\\\hspace*{0.2cm}(\select c\_name AS name, c\_phone AS phone\\
    \hspace*{0.5cm}\from customer c 
    LEFT JOIN orders o {\sc On} c.c\_custkey = o.o\_custkey\\
    \hspace*{0.5cm}\where (c.c\_acctbal <= 10000.00 OR o.o\_orderkey {\sc Is} {\sc Null})) \\
\unionall\\
\hspace*{0.2cm}(\select s\_name AS name, s\_phone AS phone\\
    \hspace*{0.5cm}\from supplier s
    \where s.s\_suppkey {\sqlin} (\\
    	\hspace*{0.5cm}\hspace*{0.5cm}\select l\_suppkey \from 
    	lineitem l JOIN orders o\\ \hspace*{0.5cm}\hspace*{0.5cm}{\sc On} l.l\_orderkey = o.o\_orderkey	\where s.s\_acctbal <= o.o\_totalprice \\
    	\hspace*{0.5cm}
      	\sqland l.l\_commitdate = l.l\_receiptdate 
      	\sqland l.l\_shipmode {\sqlin} (`AIR',`TRUCK')))\\
) as customer\_supplier
{\groupby} name, phone\\
    \end{tabular}
}
\caption{Final Synthesized Query \QE}
     \label{fig:qe_xpose}
 \end{subfigure}
\caption{Example Hidden Query Extraction with {\utwo}}
\end{figure}

\subsection{Mutation Framework in XRE}
\label{sect:mututwo}
We now explain how the mutation framework outlined in Section~\ref{sect:background} is adapted to support the extended scope of \utwo.

\subsubsection{From Clause Extractor}
\label{sec:fce}
The EbE (Extraction-by-Error) strategy is sufficient for extracting {\tH}, the set of tables in \QH. However, for Unions, we need this information at the granularity of \emph{sub-queries}. This requires bringing the {\bf EbV (Extraction-by-Voiding)} strategy also into play, as described later in Section~\ref{sect:union}.

\subsubsection{FIT-results}\label{sect:fit}
In \utwo, we consider the possibility of \QH having outer joins. Due to their presence,
attributes in the result set {\fullresult} may have NULL values even if the original input database \D is completely NULL-free. This has repercussions for our extraction procedure -- for instance, the binary search for arithmetic predicates is no longer viable. Therefore, we need to make
the definition of a populated result more nuanced than simply checking for the presence of tuples in the output.

Specifically, we introduce the notion of a {\bf fully instantiated tuple (FIT)}, a tuple not containing any NULL values. With this notion, a FIT-result is a query result that features \emph{at least one} FIT row, whereas a UNFIT-result has no such rows. 
A FIT result implies that the input database could provide one or more pairs of matching tuples, and such tuples become candidates for the join predicate extraction 
(Section \ref{sect:eq_algo}).
Consequently, for \QH to be extractable, \RH must be a FIT-result -- our modified minimizer and the filter predicate extractor ensure retention of this characteristic in the recursive halving and binary search procedure, respectively.

\subsubsection{S-Value Extractor}\label{sect:sve}
When \QH is restricted to arithmetic predicates, the s-value bounds correspond, as discussed in Section~\ref{sect:sv}, to the predicate constants in the query, i.e. they are \emph{static}.  However, this is no longer the case if \QH features algebraic predicates such as $col_x \leq col_y$. Consider \done.$col_x = X$, for some $X$ $\in [i_{min}, i_{max}]$. Then, $col_y$ has s-value interval $[X, i_{max}]$, a {\em floating interval} due to its dependency on $X$. Such floating dependencies are identified, as detailed in Section~\ref{sect:aoa},
by mutating attribute values in \dmin and \emph{iteratively} applying the  static s-value extractor used for arithmetic predicates. We denote this modified s-value extractor by {\textbf{\filterex}}.

\subsubsection{Reentrant Pipeline}
The XRE module features a generalized mutation pipeline, with multiple outgoing edges from a node, as well as looping among the extraction modules. This re-entrant structure facilitates progressive construction of \QS, the seed query.

\subsubsection{Seed Query Output}
With the above augmented mutation framework, the \QS output by XRE is shown in Figure~\ref{fig:xre_case}. It captures the \unionall, \from and \where clause predicates correctly.

\subsection{LLM-based Synthesis Framework in XFE}\label{sect:xfe_overall}

Since XRE's extraction technique is based on the single-row \done, it only extracts a minimal conjunctive flat query in \QS.
We now need to add the nested structure, outer join, and disjunctive \sqlin operator, and pull up the common grouping attributes from the sub-queries to the outer query.

XFE infers the nested structure using the LLM on the business description \tq. For instance, the phrase ``any order" in \tq (Figure~\ref{fig:utwoscope}) maps to the {\sqlin} operator, and the nested structure is inferred from it. 
As elaborated later in Section~\ref{sect:llm}, XFE has a set of guidelines for such query synthesis through automated iterative prompting that ensures eventual convergence. An important subset of the guidelines instructs the LLM to remain aligned to \QS while performing its synthesis. For instance, in the above example, retaining the join, comparison and disjunctive predicates is crucial. Also, specific guidelines instruct the LLM on how to handle result mismatches between \QS and \QH on \D. For example, moving the {\groupby} attributes to the outer query is driven by the guidelines.
 
Finally, \QE, the extracted query post-XFE intervention, is shown in Figure~\ref{fig:qe_xpose}, and it is clearly semantically identical to \QH.

\medskip
In the following sections, we cover the internals of the Union and Algebraic Predicate extractors, as well as the XFE module. Due to space limitations, we defer the Disjunction extractor to \cite{xpose_tech_report}.

%% file: union.tex
\section{UNION ALL Extraction}\label{sect:union}
The key step of union extraction is to isolate subqueries in terms of the tables in their respective \from clauses. This isolation is achieved through a fine-grained {\bf EbV} (Extraction-by-Voiding)-based scheme, described below.

\subsection{Problem Formulation}\label{union_problem}
Let $q_i$ denote a flat SPJGAOL query. A hidden union query \QH is a union of an unknown number, $n$ ($n \geq 2$), of $q_i$ subqueries, that is, \QH = $\bigcup_{i=1}^n q_i$. Let \fromtab{$q_i$} denote the set of tables present in subquery $q_i$. Then, {\tH} = $\bigcup_{i=1}^n$\fromtab{$q_i$}. 

The tables that appear in {\em all subqueries} are referred to as \comtab{\QH}, the set of \emph{common tables}. With this, {\tH} $-$ \comtab{\QH} represents the set of tables that appear only in some, but not all, subqueries in {\QH} -- this set is referred to as {\em auxiliary tables}, denoted \parttab{\QH}. 
Further, the auxiliary tables on a per-subquery basis are written as \parttab{$q_i$}.

We can now specify \fromtab{$q_i$} = \parttab{$q_i$} $\bigcup$ \comtab{\QH}.
Therefore, to uniquely extract  \fromtab{$q_i$}, which is our objective, the sets \parttab{$q_i$} and \comtab{\QH} need to be determined. In doing so, we make the following assumptions:  (1) Each subquery $q_i$ independently produces a FIT-result on \D.  (2) $\forall i, j$, $i \neq j$, \parttab{$q_i$} $\not\subset$ \parttab{$q_j$}. This assumption is usually satisfied in practice -- for instance, data integration queries compute unions over fact tables from different locations.

\subsection{Extraction Process Overview}

 \subsubsection{Compute Common Tables}
 \label{sect:computetabs}

    First, {\tH} is extracted using the EbE procedure (Section ~\ref{sect:from}).
    Then, \comtab{\QH} is identified using the EbV techninque
    in the following manner:
    A table $T$ is made {\em void}, and then \QH is executed. An UNFIT-result implies that none of the subqueries produced a FIT-result. 
    A union of conjunctive queries can produce an UNFIT-result due to a void table $T$ only if the absence of data in $T$ prevents {\em every} subquery from obtaining {\em any satisfying tuple}. 
    Therefore, $T$ is present in all subqueries. The common tables are identified by iterating this technique over all the tables in \tH.

 \subsubsection{Compute Auxiliary Tables ($q_i$)}  
 \parttab{\QH} is given by {\tH} \(-\) \comtab{\QH}.
Our goal now is to choose subsets from \parttab{\QH} so that each subset corresponds to a subquery of \QH, ensuring that every subquery has a matching subset.  
For this, we enumerate the power set of \parttab{\QH} and test each member (barring the null set and the entire set) with regard to whether it maps to a {\em single subquery}. The test is discussed next,  capturing its logic in
Algorithm \ref{algo:basic}.

\begin{algorithm}[h!bt]
\DontPrintSemicolon
\setstretch{0.85}
\small
\Input{\QH, \D, \parttab{\QH}, \comtab{\QH}}

 {\coretables} $\gets$ \parttab{\QH}

 {\sidetables}, Max-{\sidetables}, {\aux}, $FromSet$ $\gets$ $\emptyset$

$U \gets$ PowerSet(\parttab{\QH}) $-$
$\{\emptyset, $ \parttab{\QH}$\}$ \tcp{exclude $\top$ and $\bot$ from the lattice}

$U_{asc} \gets$ Sequence of sets in $U$ in increasing size\nllabel{opt_start}

\For{u $\in$ $U_{asc}$}{
    
    \If{a subset of $u$ is already in {\coretables}}{
    include $u$ in {\coretables} 
    
    {\bf continue\nllabel{opt_end}} \tcp{skip checking for $u$}
    }
    $D' \gets$ void all tables in $u$ in \D
    \nllabel{to_revert_u}

    \lIf{\QH($D'$) is UNFIT}{
        include $u$ in {\coretables}
    }
    
    Revert Mutations done by Line \ref{to_revert_u}
    }

    {\sidetables} $\gets U -$ {\coretables}

 Max-{\sidetables} $\gets$ $\{c | c \in$ {\sidetables}, $\forall t$ $\in$ {\parttab{\QH}} - \{$c$\}, $c \cup \{t\} \in$ {\coretables}\} 

\For{each $c \in$ Max-{\sidetables}}{

    include set \parttab{\QH} $-$ $c$ in {\aux}

    }

\For{each $p$ $\in$ {\aux}}{
$fromSq_i$ $\gets$ $p$ $\cup$ \comtab{\QH}

include $fromSq_i$ in $FromSet$
}
\Return $FromSet$

\caption{Union Extraction Algorithm:\\ {\em Assignment of Tables to Subqueries}}
\label{algo:basic}
\end{algorithm}

\subsubsection{Union Extraction: Table Assignment Algorithm} 
With the common tables intact, a void auxiliary table fails to satisfy the related predicates in the corresponding subqueries. 
Thus, \parttab{$q_i$} is a set of tables from \parttab{\QH}, any one of which when voided, gets $q_i$ to produce a UNFIT-result. 
To identify such sets, we consider the power set of \parttab{\QH}, and observe the result of \QH upon voiding the tables from the enumeration. 
 The algorithm first takes a member set $u$ from the power set and {\em voids} all the tables in it. If \QH produces an UNFIT-result, $u$ is included in collection {\coretables}. The set {\coretables} represents all the {\em voided states} of the database where no subquery is satisfied. Next, we get the set {\sidetables}, where voiding each member set still causes \QH to produce a FIT-result. They capture the {\em voided} database states where at least one subquery is satisfied. 

 \begin{table}
 \centering
\footnotesize
\begin{tabular}{|p{4cm}p{4cm}|}
\hline
\multicolumn{2}{|p{8cm}|}{
\QH = (\select $\dots$ \from $orders, customer$  \where $\dots$)
$\cup$ (\select $\dots$ \from $supplier$ \where $\dots$ (\select $\dots$ \from $lineitem$, $orders$ \where $\dots$))}\\\hline\hline
\multicolumn{2}{|p{8cm}|}{{\tH} = \{{\sc Orders}(o), {\sc Supplier}(s), {\sc Customer}(c), {\sc Lineitem}(l)\} 
}\\
\comtab{\QH} = $\{o\}$
&
\parttab{\QH} = $\{s, c, l\}$
\\
\multicolumn{2}{|p{8cm}|}{$U_{desc}$ = ($\{s, c\}$, $\{s, l\}$, $\{c, l\}$, $\{s\}$, $\{c\}$, $\{l\}$)
}\\
{\coretables} = \{$\{s, l, c\}$, $\{s, c\}$, $\{l, c\}$\}
&
{\sidetables} = \{ $\{s, l\}$, $\{s\}$, $\{c\}$, $\{l\}$\}
\\
Max-{\sidetables} = $\{\{s, l\}$ $, \{c\}\}$ 
&
{\aux} = $\{\{c\}$ $, \{s, l\}\}$\\
\multicolumn{2}{|c|}{$Froms$ = $\{\{c, o\}$ $, \{s, l, o\}\}$
}\\\hline\hline
\multicolumn{2}{|p{8cm}|}{\fromtab{$q_1$} = \{{\sc Customer}, {\sc Orders}\}, \fromtab{$q_2$} = \{{\sc Supplier}, {\sc Lineitem}, {\sc Orders}\}
}\\\hline
\end{tabular}
\caption{Table assignment (Algorithm \ref{algo:basic}) for \QH in Figure~\ref{fig:utwoscope}}
\label{tab:union_examples}
\end{table}

In 
Table \ref{tab:union_examples}, these sets are enumerated for the running example of \QH given in Figure~\ref{fig:utwoscope}. 
For instance, \{$s$, $c$\} $\in$ {\coretables} because voiding {\sc Supplier} and {\sc Customer} together discards the satisfying tuples for both the subqueries in \QH. 
On the other hand, \{$s, l$\} $\in$ {\sidetables} because voiding {\sc Supplier} and {\sc Lineitem} together discards the satisfying tuples only for the second subquery. 

For a set \(u\) in {\coretables}, its supersets are also in {\coretables}. If voiding tables in \(u\) obtain an UNFIT-result for \QH, voiding of more tables cannot produce FIT-tuples. This fact is leveraged to reduce the iterations of voiding, by checking elements from the power set lattice bottom-up.
Lines \ref{opt_start}-\ref{opt_end} of the algorithm capture it.

Next, we identify the {\em maximal members} of {\sidetables}. 
Adding any table from \parttab{\QH} to these members gets a set already in 
{\coretables}.
E.g. set \{$c$\} is in Max-{\sidetables} because when {\sc Customer} table is void, \QH produces a FIT-result, and if any other table is voided, say $l$ ({\sc Lineitem}), the result of \QH becomes UNFIT (set \{$c, l$\} belongs to {\coretables}). Thus, the construction of Max-{\sidetables} isolates the individual subqueries. 
When one member set of Max-{\sidetables} is void, exactly {\em one subquery is active}.

\begin{figure}[!h]
\centering
\includegraphics[width=0.85\columnwidth]{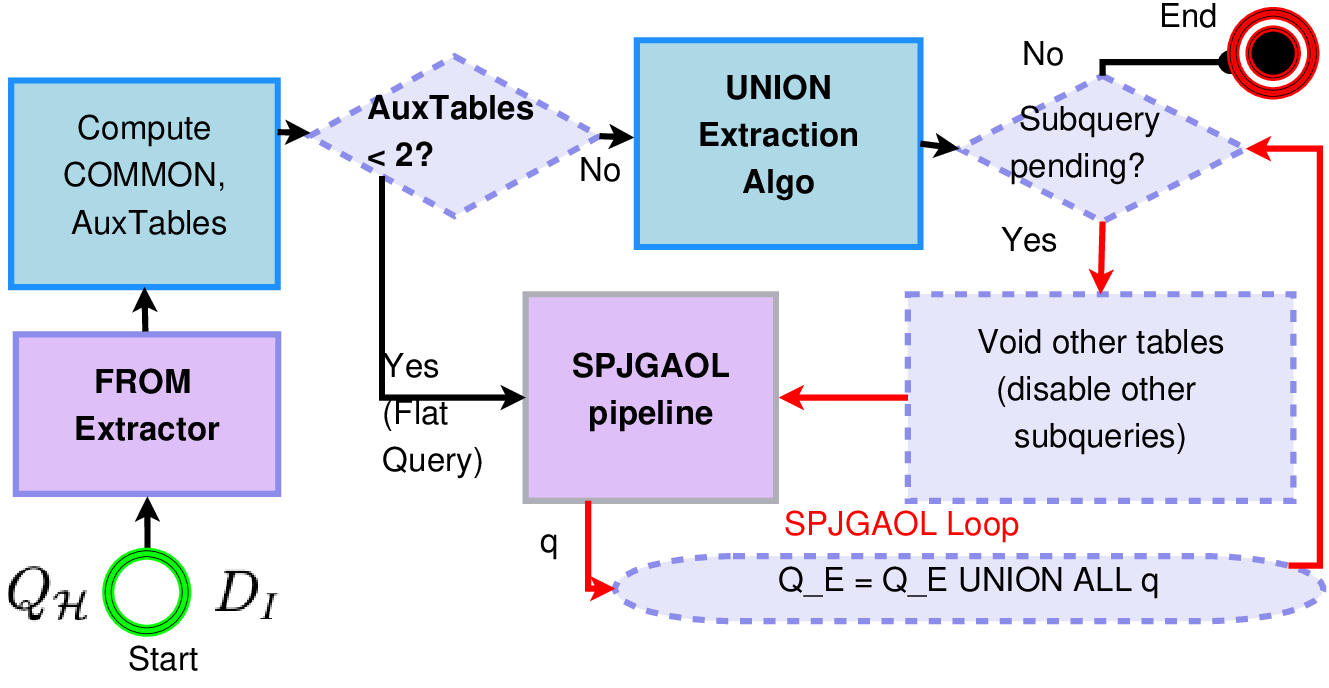}
\caption{Flow of Union Detection and Extraction}
\label{fig:pipeline}
\end{figure}

\subsubsection{Iterative Subquery Extraction} 
After \parttab{$q_i$} is computed, the subqueries of \QH are extracted in a loop (SPJGAOL loop in Figure~\ref{fig:pipeline}).
In each iteration $i$, all tables other than \fromtab{$q_i$} are voided, to ensure that execution of \QH produces results only from $q_i$. Then the \uone pipeline is used to extract the SPJGAOL-subquery.  The overall flow is shown in Figure \ref{fig:pipeline}. In this context, note that a nested subquery is extracted as a flat query by Algorithm~\ref{algo:basic} (similar to what happened for $q_2$ in Figure~\ref{fig:xre_case}). 

\subsubsection{Mutation Overheads}
Executing \QH over the power set enumerated in Algorithm~\ref{algo:basic} could, in principle, be computationally highly expensive. However, in practice, the number of auxiliary tables is often limited -- for instance, it could correspond to the fact-tables in a warehouse schema, which is usually a small value -- so even the power set may not be unviably large.  Further,  execution overheads could be reduced by various optimizations, such as constructing a rich set of column indexes, as explained in ~\cite{xpose_tech_report}.

%% file: aoa.tex
\section{Algebraic Predicates}
\label{sect:aoa}
In \utwo, we consider extraction of predicates comparing a pair of attributes (i.e. \emph{column op column}), termed as algebraic predicates (\alp) to distinguish from the arithmetic predicates which are \emph{column op value}.  The {\alp} attributes may be from the same table (i.e. a filter predicate) or different tables (i.e. a join predicate), while {\op} is a binary comparison operator from $\{=, \leq, \ge, <, > \}$. Further, this formulation supports treating many-to-many and non-equi-joins on par with key-based equi-joins, overcoming a major limitation of {\uone} where only the latter were permitted.

We now discuss how to extract algebraic predicates using database mutation from the signatures available in \done.

\subsection{Extraction Process Overview}
Algebraic predicate extraction is initiated on \done, which as mentioned before, is ensured to produce a FIT-result.
The s-value extractor {\filterex} (Section~\ref{sect:sve}) is executed to compute the s-value intervals ({\svalinterval}) of all attributes in \tH. The s-value interval for column $c$ is denoted as {\svalinterval}$_{c}$, and {\svalinterval}$_{c}^{LB}$ and {\svalinterval}$_{c}^{UB}$ are used to represent the lower and upper bounds, respectively, of this interval.

In Figure~\ref{fig:aoa_si_se}(a), a sample \done for the $q_2$ subquery of Figure~\ref{fig:utwoscope} is shown, and the {\svalinterval} obtained from this \done is listed in Figure~\ref{fig:aoa_si_se}(b) for the various attributes appearing in the query.

\begin{figure}
\centering        
{\footnotesize

\begin{tabular}{|c|c|c|c|}
\multicolumn{4}{c}{\bf (a) Minimized Database \done}\\
\multicolumn{4}{c}{Table: \done.{\sc Lineitem}}\\\hline
{$l\_orderkey$} & {$l\_commitdate$} & {$l\_receiptdate$}& $l\_suppkey$\\
\hline
2739811 & 1995-03-16 & 1995-03-16 &1793\\
\hline\end{tabular}

\begin{tabular}{|c|c|c|c|c|c|}
\multicolumn{2}{c}{Table: \done.{\sc Orders}}&\multicolumn{1}{c}{}& \multicolumn{3}{c}{Table: \done.{\sc Supplier}}\\\cline{1-2}\cline{4-6}
{$o\_orderkey$}&$o\_totalprice$ &&{$s\_acctbal$}&$s\_suppkey$ &$\dots$\\\cline{1-2}\cline{4-6}
2739811  &150971.81& &2530.46 &1793 &$\dots$\\
\cline{1-2}\cline{4-6}
\end{tabular}

}

\smallskip

{\footnotesize
\medskip
\begin{tabular}{|p{1.5cm}|p{1.8cm}|p{1.5cm}|p{2cm}|}
    \multicolumn{4}{c}{\bf (b) Extracted S-value intervals from \done}\\\hline
    {\bf Attribute} & {\bf SVI} &{\bf Attribute} & {\bf SVI} 
    \\\hline\hline
   \multirow{2}{=}{\centering $l\_orderkey$, $o\_orderkey$} & 
    \multirow{2}{=}{\centering $[2739811, 2739811]$} & 
    $s\_acctbal$ & $[i_{min}, 150971.81]$ \\\cline{3-4}

    & & $o\_totalprice$ & $[2530.46, i_{max}]$ \\\hline
        $l\_suppkey$, $s\_suppkey$& [1793,1793]&  
        $l\_commitdate$, $l\_receiptdate$& [1995-03-16, 1995-03-16]\\\hline
\end{tabular}
\smallskip

}\caption{\done and \svalinterval for $q_2$ of Figure~\ref{fig:utwoscope}}
    \label{fig:aoa_si_se}
\end{figure}

The {\alp}-extraction algorithms are executed with \done and {\svalinterval} as inputs. For reasons described later in this section, we have devised separate algorithms to extract inequality and equality predicates, described in Sections~\ref{sect:ineq_algo} and ~\ref{sect:eq_algo}, respectively.

\input{ineq}

%% file: ineq.tex
\subsection{Inequality ($\leq$, $\geq$, $<$, $>$) Predicate Extraction}
\label{sect:ineq_algo}
 
XRE enumerates all candidate pairs of inequality attributes and then systematically validates them for presence in the query. The candidates are conceptually represented in a graph, with the attributes being the vertices and an inequality $col_x$ $\leq$ $col_y$ resulting in an edge $col_x \rightarrow col_y$.

\subsubsection{\bf Enumerate Inequality Candidates}
\label{sect:makeE}
The candidate identification and graph construction is done using the 
 \done and {\svalinterval} inputs. As a concrete example of this process, 
consider Figures~\ref{fig:aoa_si_se}(a) and (b). Here, we have \done.$s\_acctbal$ = 2530.46, and \svalinterval$_{o\_totalprice}^{LB}$ also equal to 2530.46.
So, $s\_acctbal \leq o\_totalprice$ is a possibility, and therefore edge $s\_acctbal \rightarrow o\_totalprice$ is added to the graph.

\subsubsection{\bf Mutate to Float the S-value Bounds}\label{sect:bound_impact}
Now, we have to check whether $col_x \rightarrow col_y$ holds. To do that, we need to confirm whether \svalinterval$_{y}^{LB}$ is {\em floating}, that is, dependent on the current \done value of $col_x$.  
So, $col_x$ is first mutated with \svalinterval$_{x}^{UB}$, and then {\filterex} is used on $col_y$ to extract its SVI. If \svalinterval$_{y}^{LB}$ also now changes to \svalinterval$_{x}^{UB}$, we confirm $col_x \leq col_y$. Alternatively, if  \svalinterval$_{y}^{LB}$ changes to a number larger than \svalinterval$_{x}^{UB}$, then $col_x < col_y$ is confirmed.

\begin{figure}[h!bt]
  \centering
  \includegraphics[width=0.9\linewidth]{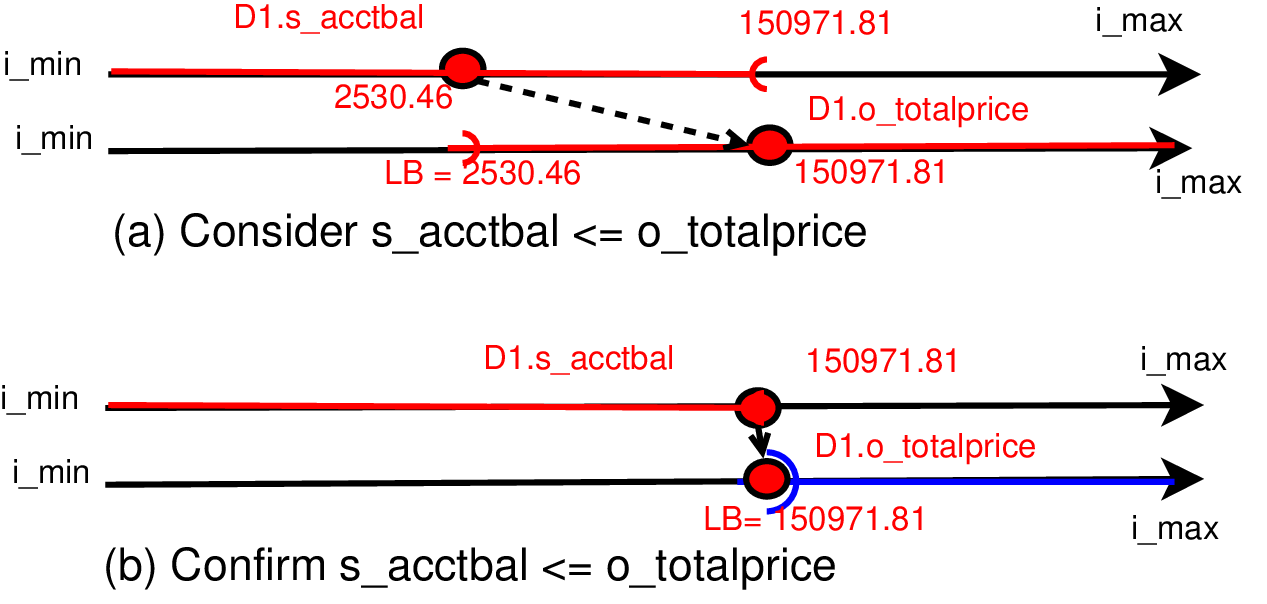}
  \caption{Identifying dependency between two attributes}
  \label{fig:aoa_dep}
\end{figure}

For the running example, 
we first represent the possibility $s\_acctbal \leq o\_totalprice$ as a dashed directed edge in   
Figure \ref{fig:aoa_dep}(a).  Then, in Figure \ref{fig:aoa_dep}(b),  \done.$s\_acctbal$ is mutated with its SVI UB of 150971.81, which causes $o\_totalprice$ to have a new SVI LB of 150971.81. Thus, $s\_acctbal$ $\leq$ $o\_totalprice$ is present in the query.

\subsection{Equality (=) Predicate Extraction}\label{sect:eq_algo}
At first glance, it might seem that the inequality extraction described above could be easily overloaded to also determine equality -- that is, 
if both $col_x \leq col_y$ and $col_y \leq col_x$ are extracted, then equality predicate $col_x = col_y$ is confirmed. However, there is a catch, as follows:  With equality,  \done.$col_x$ = \done.$col_y$ = $v$ is guaranteed, for some $v$ within the domain of $col_x$ (equivalently $col_y$). This causes {\svalinterval}$_x$ = {\svalinterval}$_y$ = [$v$, $v$] -- therefore, in contrast to inequalities, we are unable to mutate the s-values independently while retaining a FIT-result from \QH.

To get around this hurdle, mutations are concurrently performed for \emph{both attributes}. 
Returning to the example in Figure~\ref{fig:aoa_si_se}, we find {\tt l\_orderkey} = {\tt o\_orderkey} = 2739811, and {\tt l\_commitdate} = {\tt l\_receiptdate} = 1995-03-16. Therefore, these two pairs are candidates for such mutation.
A binary search-based technique, similar to Section~\ref{sect:where}, is used to extract their common s-value intervals. If the interval is {\em wider} than [$v$, $v$], implying a {\em floating interval}, algebraic equality is confirmed. In the example, both pairs are thus extracted as algebraic equality predicates. 

\subsubsection{Outer Join Predicates}\label{thm:outerjoin_dmin}
The above logic is not devised to extract outer joins, but it goes part of the way -- it identifies them as equi-join predicates. In the above mutation-based technique on \done, to produce a FIT-result, the join attributes need matching values. Mismatching values in \done cause \QH to produce UNFIT-result. So, only the matching values are identified as s-values for the attributes, and hence they are extracted as equi-joins.
As a result, the LOJ in subquery $q_1$ of Figure~\ref{fig:utwoscope} is extracted as $c\_custkey$ = $o\_custkey$, as shown in Figure~\ref{fig:xre_case}.

\subsubsection{Semi-join Semantics}\label{thm:semi_join}
For semi-joins, which evaluate the existence of \emph{any} match, rather than all matching tuples, minimizing \D to \done causes the FIT-result from \QH to depend solely on a single matching tuple. This leads to the semi-join being partially extracted as an equi-join, similar to outer join. As a consequence, the semi-join between $l\_suppkey$ and $s\_suppkey$ in Figure~\ref{fig:utwoscope} is extracted as $l\_suppkey$ = $s\_suppkey$ in Figure~\ref{fig:xre_case}.

%% file: llm.tex
\section{Forward Engineering in \utwo}\label{sect:llm}
We now turn our attention to XFE, where the predictive abilities of LLMs are used on the textual description \tq in conjunction with the grounding provided by \QS, the seed query output by XRE.

\subsection{Basic Synthesis Prompt}\label{sect:basic_prompt}
XFE initiates the synthesis task using the prompt template listed in Table ~\ref{tab:basec_prompt}(a) -- this ``initial prompt" (IP) comprises the
(i) textual description \tq, (ii) schema specification, (iii) seed query \QS produced by XRE, (iv) cardinalities of \RH and \RS, and (v) general guidelines 
on synthesizing from \QS.

The LLM typically finds it easy to infer complex constructs such as nested structures, outer joins, and existential operators, since they are usually expressed directly as such in the business description. As a case in point, TPCH `Customer Distribution Query' Q13 specifies listing customers and their orders, {\em including those who have no order} -- this clearly maps to outer join. Consequently, even though XRE extracts an outer join as an equi-join predicate (Section~\ref{thm:outerjoin_dmin}), XFE is able to refine it correctly.

\begin{figure}
\fcolorbox{black}{gray!5}{%
\small
  \begin{minipage}{0.9\linewidth}
\newcounter{itemcount} 
\subsubsection*{\bf Basic Guidelines:}\label{basic}
\begin{enumerate}[wide, labelwidth=!, labelindent=0pt,label=G\arabic*]
\setcounter{itemcount}{0} 
\item\label{legal} Do not formulate syntactically incorrect SQL.\stepcounter{itemcount} 
\item\label{repeat} Do not repeat any previously formulated incorrect SQL.\stepcounter{itemcount} 
\item\label{specific_1} Do not use redundant join conditions or redundant nesting. \stepcounter{itemcount} 
\item\label{placeholder} Do not use any predicates with place holder parameters.\stepcounter{itemcount} 
    \end{enumerate}

\subsubsection*{\bf Guidelines to align synthesis with \QS:}\label{seed_align}
\begin{enumerate}[wide, labelwidth=!, labelindent=0pt,label=G\arabic*]
\setcounter{enumi}{\theitemcount}
    \item\label{alltables} Strictly use the tables given in \QS. \stepcounter{itemcount} 
    \item\label{tableinstances} If \QS has a multi-instance table in its \from clause, keep all the table instances in your query.\stepcounter{itemcount} 
    \item\label{notjoin} Do not use join predicates absent from \QS.\stepcounter{itemcount} 
\item\label{projection} Strictly reuse the order, attribute dependencies, and aliases of the projections from \QS.\stepcounter{itemcount} 
\end{enumerate}

\subsubsection*{\bf Guidelines to align synthesis with \tq:}\label{seed}
\begin{enumerate}[wide, labelwidth=!, labelindent=0pt,label=G\arabic*]
\setcounter{enumi}{\theitemcount}

    \item\label{filters} Validate all the predicates in the seed query against \tq. Include all the valid predicates in your query.\stepcounter{itemcount} 
    \item\label{invalid} For the attributes in the invalid filter predicates, validate their use from \tq.\stepcounter{itemcount} 
    \item\label{joinvsin} A semi-join, implying at least one match, maybe incorrectly present as an equi-join in \QS.\stepcounter{itemcount} 
\end{enumerate}

\subsubsection*{\bf Guidelines to synthesize compact and meaningful queries:}\label{perf}
\begin{enumerate}[wide, labelwidth=!, labelindent=0pt,label=G\arabic*]
\setcounter{enumi}{\theitemcount}
    \item\label{cte} A subquery used more than once should be a CTE with alias.\stepcounter{itemcount}

    \item\label{one_count} A subquery may have at most one COUNT() aggregation.\stepcounter{itemcount} 
\end{enumerate}

\subsubsection*{\bf Guidelines to address result mismatch:}\label{result}
\begin{enumerate}[wide, labelwidth=!, labelindent=0pt,label=G\arabic*]
\setcounter{enumi}{\theitemcount}
\item\label{groupby_union} If \RS has more rows than \RH, consider performing \unionall before \groupby.\stepcounter{itemcount} 
    \item\label{nested_groupby}
    If \RS has fewer rows as compared to \RH,  consider either adding more {\groupby} attributes or having more {\groupby} clauses through nestings.\stepcounter{itemcount} 
\end{enumerate}
\end{minipage}
}
\caption{Guidelines for Query Synthesis by XFE}
\label{fig:guidelines}
\end{figure}

However, for several other operators, including semi-joins and membership operators, the LLM has a tendency to occasionally go off the rails and produce spurious constructs. To minimize this possibility, we include the detailed guidelines shown in Figure~\ref{fig:guidelines} which put in explicit guardrails to make the LLM produce relevant SQL.  The guidelines range from the obvious (\ref{legal}: ``Do not formulate syntactically incorrect SQL'') to compliance with \QS for its provably correct aspects (\ref{alltables}: ``Strictly use only the tables given in the seed query'') to more subtle aspects such as not having multiple Count aggregations (\ref{one_count}: ``A subquery may have at most one COUNT() aggregation.'') and checking the validity of \QS predicates (\ref{filters}: ``Validate all the predicates in the seed query against the textual descriptions.'').  We have found these guidelines sufficient to handle the queries investigated in our study, but it is, of course, possible that a few more may have to be added for other scenarios.

For the example \QS in Figure~\ref{fig:xre_case}, \ref{joinvsin} brings in the nested structure and IN operator to rewrite the $l\_suppkey$ = $s\_suppkey$ of \QS into a semi-join. Further, moving the {\groupby} operator to the outer query is triggered by \ref{groupby_union}. In this manner, the prompts direct the LLM towards an accurate extraction.

\begin{table}[h!bt]
    \centering
    {\small
    \begin{tabular}{|p{8cm}|}
    \hline
    \cellcolor{gray!20}{\bf \em (a) Initial Prompt [IP]}\\
         You are an expert in formulating SQL queries from high-level textual business descriptions. \\
         Formulate SQL query for the following description:{\color{blue}<\tq>}\\
        Use the following schema to formulate SQL:{\color{blue}<Schema DDL>}\\
        Use the following SQL as a seed query. You should refine the seed query to produce the final SQL:{\color{blue}<\QS from XRE>}\\
        Follow the refinement guidelines mentioned below:{\color{blue}<Guidelines>}\\\hline
        \hline
   \cellcolor{gray!20}{
         \bf\em (b) Result-Correction Prompt [RCP] \newline (Query aligned with \QS, but result does not match \QH)}\\
         \cellcolor{lightviolet}{\bf [RCP.v1]}\\
        You formulated the following SQL:{\color{blue}<Last returned SQL \QE>}\\
         It produces the following number of rows:{\color{blue}|<\QE Result>|}\\
         Below is the actual result cardinality:{\color{blue}|<\QH Result>|}\\
         The results do not match. Fix the query.\\
         \cellcolor{lightviolet}{\bf [RCP.v2]}\\
        You formulated the following SQL:{\color{blue}<Last returned SQL \QE>}\\
         Its result does not match with actual result. Fix the query.\\\hline\hline
         \cellcolor{gray!20}{
        \cellcolor{gray!20} \bf\em (c) Clause-Correction Prompt [CCP] \newline (Query is syntactically incorrect or misaligned with \QS)}\\
        You formulated the following SQL:{\color{blue}<Last return SQL \QE>}\\
         Fix its {\color{blue}<incorrect clause>} as per \QS (repeat for relevant clauses)\\\hline
    \end{tabular}
    }
    \caption{Query Synthesis Prompts in XFE}
    \label{tab:basec_prompt}
\end{table}

\subsection{Error Feedback Prompts}
Feedback prompting is triggered when the synthesized query from the initial prompt differs wrt \QH in terms of the results on \D. Specifically, a ``result-correction'' prompt (RCP), shown in Table~\ref{tab:basec_prompt}(b), is submitted to the LLM asking for query correction to address the mismatch. A mismatch in the results may be due to a difference in cardinality itself, or a difference in content even with the same cardinality. Two alternate RCPs are designed (v1 and v2, in the figure) to handle the respective cases.

Apart from result mismatch, it is also possible that the synthesis may introduce new elements not compliant with the provably correct components of \QS.  To address such errors, a ``clause-correction'' prompt (CCP), shown in Table~\ref{tab:basec_prompt}(c) is submitted to the LLM. 

\begin{figure}[!h]
    \centering
    \includegraphics[width=0.8\linewidth]{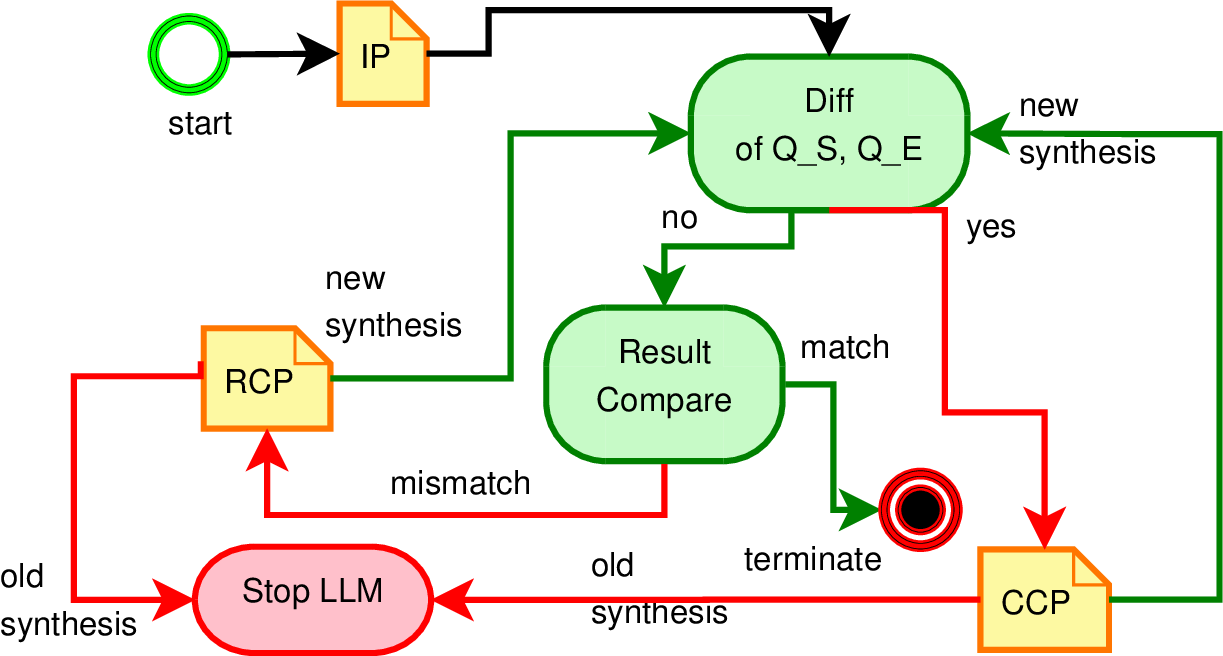}
    \caption{Automated Prompting Flow in XFE}
    \label{fig:llm_flow}
\end{figure}

These feedback prompts are iteratively exercised.  The overall prompting flow in XFE is shown in Figure~\ref{fig:llm_flow}. Prompting is repeated until one of the following occurs: (1) the results match, signaling a successful termination; (2) the results of the synthesized query do not change wrt the previous incorrect formulations. 
However, it is possible that, despite several feedback prompt repetitions, we fail to achieve a satisfactory extraction. 
For such cases (which only occurred rarely in our evaluation in Section~\ref{sect:expt}), the prompting cycle is terminated due to exceeding a threshold number of unsuccessful trials (6 in our case), and then 
we carry out an enumerative combinatorial synthesis, described next.

\subsection{Combinatorial Synthesis of Nested  Clauses}\label{sect:combinatorial}
XFE keeps the previously synthesized nesting structure constant while carrying out combinatorial synthesis. It tries valid re-distributions of clause elements between the outer and inner queries one by one, until a successful outcome is reached or the candidate pool is exhausted, signaling extraction failure.  In particular, the following two re-distributions are attempted within the nested structure: First, the tables in the outer and inner \from clauses are redistributed, along with their associated predicates, in each synthesized candidate.  Second, we know that XRE correctly identifies all the {\groupby} attributes from the {\em base tables}. Further, that two layers of {\groupby} is reasonable only when the outer layer is formed by the projections of the inner subquery. Based on this fact, we enumerate all possible candidates for outer \groupby, \select, and \orderby clauses. 
These redistribution strategies are detailed, with examples, in  \cite{xpose_tech_report}.

Depending on the query complexity, the computational overheads of the above trial-and-error exercise could, in principle, be highly expensive. However, in our experiments, we found that a successful outcome was reached within a few iterations.

\subsection{Equivalence Checker} 
\label{sect:verify}
The XRE-XFE pipeline may end with either the appearance of a successful extraction, or an explicit failure. In the former case, there still is the possibility of a \emph{false positive}, where equivalence between the extracted query and the hidden query is incorrectly claimed. Therefore, we need to implement checks to either eliminate or, at least, reduce such possibilities.  At first glance, an obvious verification mechanism is to use \emph{logic-based} query equivalence tools (e.g. QED~\cite{qed}, SQLSolver~\cite{sqlsolver}) to compare \QH and {\QE} -- the problem here is that \QH is not available in our framework. 

A more practical alternative is to use \emph{data-based} equivalence tools such as XData~\cite{xdata}, where carefully curated databases are created that elicit differences in the results between an ``instructor version'' (in our case, \QE) and a potentially incorrect ``student version'' (in our case, \QH).  However, such tools have limited scope as yet -- for instance, they do not handle scalar functions.

A final option is to use \emph{result-based} equivalence tests where several randomized databases are created on which \QH and \QE are run. The results are compared via set difference, and a non-zero outcome indicates an extraction error. Of course, result-based equivalence is only probabilistic, and not deterministic, wrt the validity of its outcomes.

%% file: expt.tex
\section{Experimental Results}\label{sect:expt}
In this section, we quantitatively evaluate the extraction performance of {\utwo}. The experiments are carried out on the \pg v14 database engine, hosted on an Intel(R) Core(TM) i9-7900X CPU @ 3.30GHz, 32 GB 2666 MHz DDR4, Ubuntu 22.04 LTS platform. The popular
GPT-4o~\cite{chatgpt_citation} LLM, configured with 0 temperature to minimize hallucinations, is the synthesis agent in the XFE module.

We present results for the following three complex query suites:
\begin{description}[leftmargin=7pt,labelindent=0pt]
\item[1. {\tpch}:] The standard decision support benchmark~\cite{tpch}, which models a data warehousing environment and features 22 analytic queries labeled Q1 through Q22.  The business descriptions are taken from the official documentation~\cite{tpch}.
\item[2. {\etpch}:] Only key-based equi-joins are modeled in {\tpch}, and there are no Unions of sub-queries. However, as highlighted in \cite{dsb,union_integration}, many-to-many joins, non-equi-joins,  and unions are commonplace in contemporary applications such as data mediators and integrators. Therefore, we have extended the basic {\tpch} schema and its query suite as follows: Union is included in some of the existing queries (by replacing {\sc Lineitem} with {\sc Web\_Lineitem} and {\sc Store\_Lineitem} tables, representing data from online and offline retail, respectively), and non-key-based joins are brought in via two new queries, Q23 and Q24, similar to those created in \cite{dsb} -- the full details are in \cite{xpose_tech_report}.  This extended benchmark is referred to as {\etpch}, and its textual inputs were created by mildly augmenting the corresponding {\tpch} descriptions. 
\item[3. STACK:] A real-world benchmark from StackExchange~\cite{marcus_stack_dataset} with 16 query templates representing questions and answers from experts. A random instance of each template is taken. Since the benchmark does not provide textual summaries, we used an LLM to create draft versions and then manually refined the descriptions. 
\end{description}
For ease of presentation, we focus here only on the benchmark queries that are 
not fully extractable by XRE or the LLM in isolation -- that is, where both modules of \utwo had to work \emph{together} to produce a successful extraction. The number of such ``bi-directional'' queries is {\bf 13} for {\tpch}, {\bf 23} for {\etpch}, and {\bf 4} for STACK.

\subsection{{\tpch} Extraction}\label{sect:expt_tpch}
We begin by executing \utwo on encrypted (via a plugin) versions of the 13 {\tpch} queries. Apart from our own manual verification, the accuracy of each extraction was also checked against the techniques discussed in Section~\ref{sect:verify}, and these results are shown in Table~\ref{tab:prove}. We observe that 5 queries -- {\bf Q2, Q13, Q16, Q20, Q21} -- could be successfully verified by {\xdata}~\cite{xdata}, our best choice from a deployment perspective. 
For queries that were outside its scope, the 
Result-equivalence-based techniques were invoked and no false positives or negatives were observed. Finally, as a matter of abundant caution, we also used the logic-based tools, SQL Solver and QED, for queries in their coverage (of course, as mentioned before, these tools cannot be used in deployment due to non-availability of hidden query). We see that QED and SQL Solver additionally confirm {\bf Q18} and {\bf Q22}, respectively, beyond those verified by {\xdata}.

\begin{table}[!h]
    \centering
    {\small
    \begin{tabular}{|c|l|}
    \hline
    {\em Equivalence Checker} & {\em {\tpch} Query ID}\\\hline\hline
    {\xdata}~\cite{xdata}& 2, 13, 16, 20, 21\\\hline
      QED \cite{qed} & 18, 21 \\\hline
    SQLSolver~\cite{sqlsolver}&2, 21, 22 \\\hline
    \end{tabular}
    }
    \caption{Checkers for Extraction Accuracy (\tpch)}
    \label{tab:prove}
\end{table}

\subsubsection{Extraction Overheads}
\begin{figure}[!h]

    \begin{tabular}{c}
    {\includegraphics[width=\linewidth]{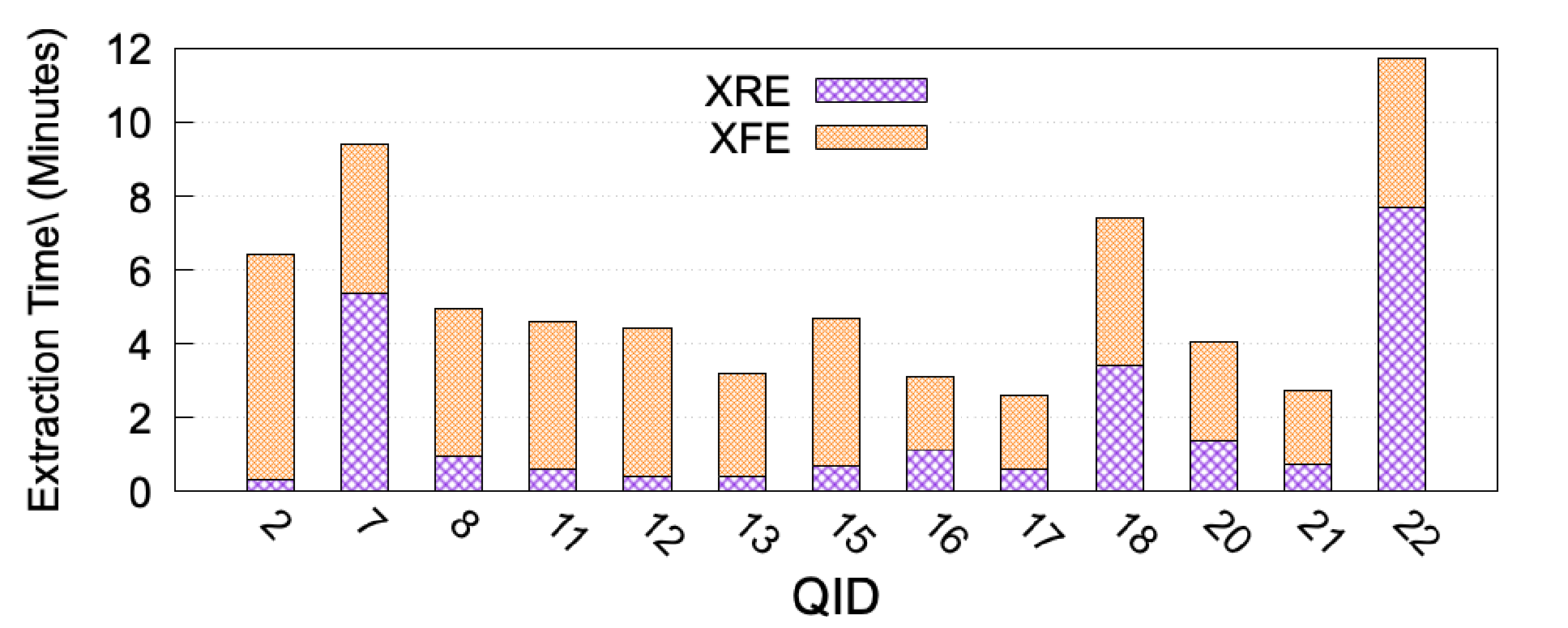}} \\
    (a) Extraction Times\\
        {\includegraphics[width=0.95\linewidth]{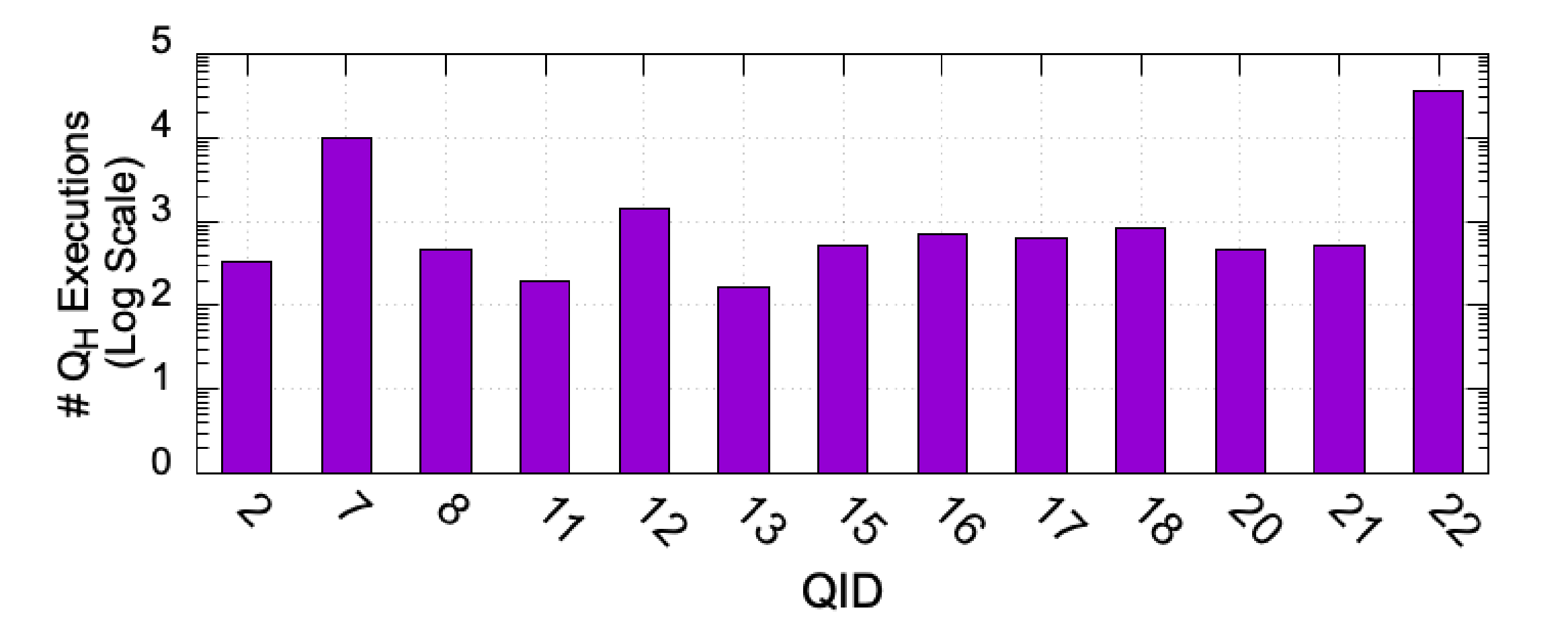}} \\
(b)  \QH Invocations
         \\
    \end{tabular}
\caption{Extraction Overheads ({\tpch})}
\label{fig:tpch}
\end{figure}

We now turn our attention to the time overheads incurred by the extractions.
These results are shown in Figure~\ref{fig:tpch}(a)  and we find that the extractions are typically completed in less than \emph{ten minutes} -- the sole exception is Q22, which has several disjunctions (7 each in two tables), and even this ``hard-nut'' query is drawn out in about 12 minutes. These overheads appear reasonable given that HQE is expected to be typically invoked in an \emph{offline} environment.

The graph also shows the time-split across the XRE and XFE modules, and we find that the distribution is query-specific -- some queries (e.g. Q7, Q22) have XRE dominating, whereas in others (e.g. Q15), XFE takes the lion's share.

The larger duration of XRE is caused by disjunction predicates, especially those with many string literals, such as in Q22. Extracting each literal requires one round of database minimization, shooting up the extraction time. On the other hand, XFE takes a longer time than XRE when the nesting structure of the query is complex. For instance, Q15 has two levels of nesting, which required multiple synthesis trials upon result mismatch.

\subsubsection{Application Invocations}
As mentioned previously, \utwo is based on generating a curated series of input-output examples by repeated invocations of the opaque executable. To quantify this notion, the number of invocations are shown (on a log-scale) in Figure~\ref{fig:tpch}(b). We observe here that most of the queries take \emph{several hundred} invocations, and a few (Q7, Q12 and Q22) go well beyond even this mark -- in fact, Q22 is more than \emph{ten thousand}!

From a conceptual perspective, these results demonstrate that (a) HQE for industrial-strength queries is a challenging problem, requiring numerous examples to achieve the goal of precision extraction, but (b) thanks to database minimization, the overheads are kept in practical check despite the numerous invocations. 

\subsubsection{Synonymized Databases}
A pertinent question that could be asked here is whether the good extraction performance is an \emph{artifact} of GPT-4o being previously trained on \tpch, which is publicly available. To assess this concern, we created a {\em synonymized} version of the benchmark (we choose synonymization rather than anonymization to ensure that the business descriptions continue to retain meaning). Specifically, the names of the tables and attributes were renamed using English synonyms, as well as synonyms from other languages. Correspondingly, the texts were also edited with the synonyms (the whole sentence remains in English). 

The good news is that, despite these material changes, all the {\tpch} queries continued to be extracted correctly. However, they required a couple more iterations of the clause-correction prompts to reach extraction closure.

\subsubsection{Choice of LLM} A second relevant question is whether GPT-4o could have been substituted with a smaller model. We tried out a variety of alternatives, including Llama 3.2~\cite{touvron2023llama}, Qwen 2.5 (both decoder-only and coder versions)~\cite{qwen25}, and DeepSeek-r1~\cite{deepseekai2025deepseekr1} -- but none of them were able to extract \emph{any} of the queries evaluated in our study. These results again suggest that industrial-strength HQE is a complex learning task requiring powerful models and reasoning power.

\subsection{{\etpch} Extraction}\label{sect:expt_etpch}

We now turn our attention to the {\etpch} query suite which features unions of sub-queries, additional nesting in the \from clause, and a rich set of join types. 
Here, 7 extraction cases (the same queries reported in Table~\ref{tab:prove}) were in the scope of {\xdata} and the formal checkers, which confirmed their equivalence. For the remaining  
17 ``bi-directional" queries, apart from our manual verification, result-equivalence tests confirmed extraction accuracy.

\subsubsection{Extraction Overheads}
 
\begin{figure}[h!bt]
\begin{tabular}{c}
     {\includegraphics[width=\linewidth]{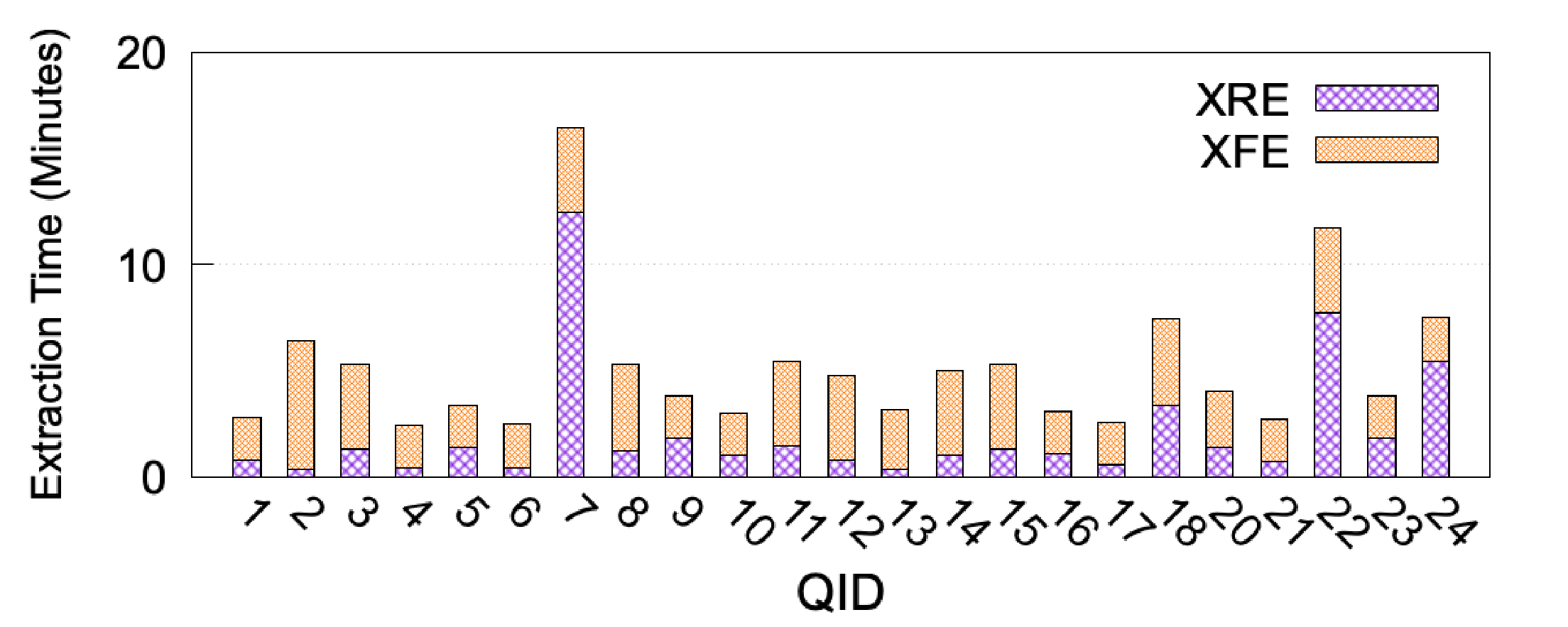}} \\
     (a) Extraction Times\\
     {\includegraphics[width=0.95\linewidth]{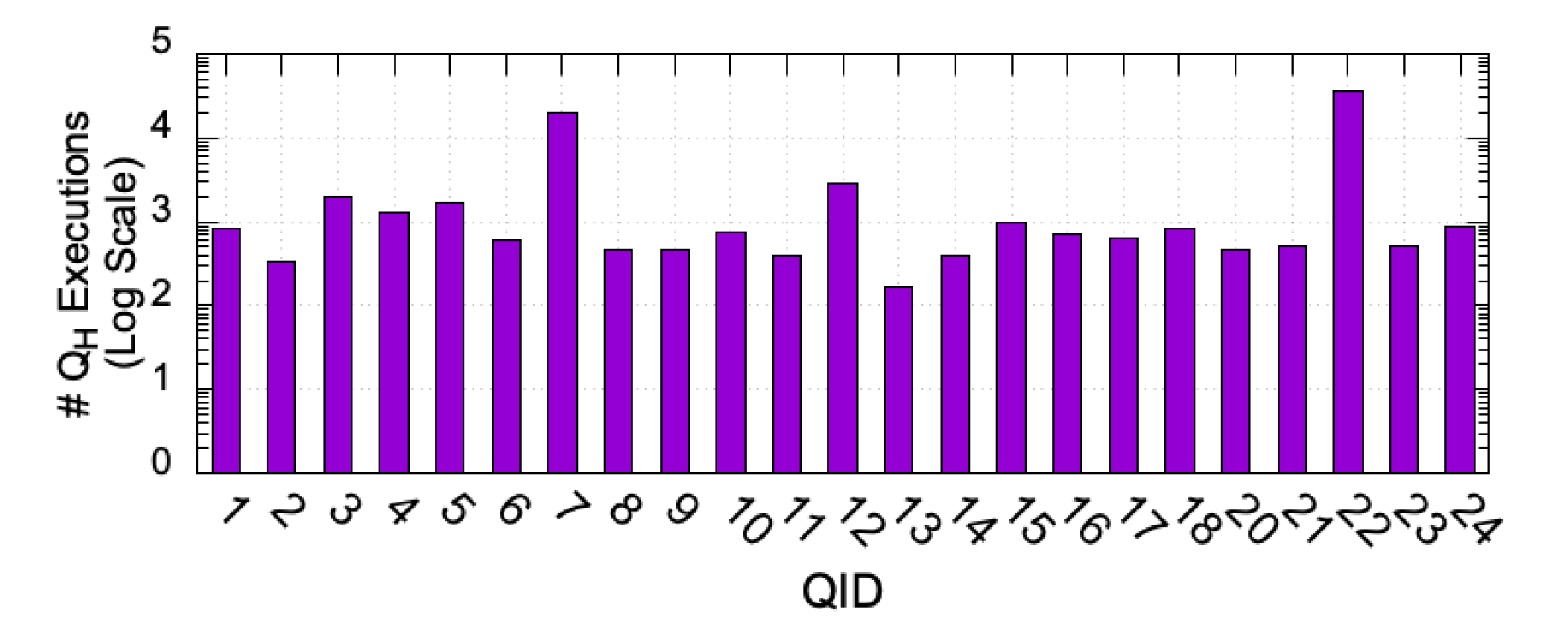}} \\
     (b) \QH Invocations\\
\end{tabular}
\caption{Extraction Overheads ({\etpch})}
\label{fig:extract_overhead}
\end{figure}

The extraction times for the {\etpch} queries are shown in Figure~\ref{fig:extract_overhead}a. We observe that most of the queries are extracted within 5 minutes, while the remaining few are completed within 15 minutes. Again, as in {\tpch}, Q7 and Q22 take the maximum time to execute \QH due to several string disjunctions and multiple instances of tables. We also observe that the time-split ratio between XRE and XFE is query-specific and covers a large range of values. Finally, with regard to invocations, shown in Figure~\ref{fig:extract_overhead}b, we see here too that they are in the several hundreds to thousands, with most being higher than their corresponding avatars in {\tpch} due to the increased query complexities.

\subsection{Drill-down Analysis ({\etpch})}
The {\etpch} scenario foregrounds the need for bi-directional engineering, as shown in Table~\ref{tab:seed_correct}, where the work done by XRE and XFE is shown on a clause-by-clause basis. We observe that while XRE does do the majority of the work, XFE also plays a significant role in taking the extractions to closure.

\begin{table}[!h]
    \centering
    {\small
    \begin{tabular}{|l|c|c|}\hline
   {\em Clause/Operator Type} & {\em XRE}&{\em XFE}\\\hline\hline
   $T_E$  (23)    & 23 &0\\\hline
  \unionall (11) & 11 &0 \\\hline
Semantic Preserving  Join Predicates (21) & 16 & 5 \\\hline
  Algebraic Inequality Predicates (5) & 3 & 2 \\\hline
  Disjunctive attributes and literals (4)&  4 &0 \\\hline
 Projection Dependencies (23) & 14 & 9 \\\hline
Semantic Preserving  \groupby (15)& 3&12\\\hline
    \end{tabular}
    }
    \caption{Extraction Distribution of XRE and XFE (E-TPCH)}
    \label{tab:seed_correct}
\end{table}

\subsubsection{XFE Prompts}\label{sect:expt_xfe}
Drilling down into XFE, the prompt sequences leading to successful extraction are shown in Table~\ref{tab:prompt_count} on a per-query basis, along with the overall token counts.
We find that all of the 20 queries shown in rows 1 and 2 of the table, are completed within 4 prompts. 
In the remaining 3 queries, Q20 required additional clause-corrections, whereas Q2 and Q13 proved to be ``feedback-prompt-resistant'' after nesting structures were synthesized by the initial prompt, requiring invocation of the computationally heavy Combinatorial Synthesis step as a last resort for extraction closure -- specifically, Q2 required redistribution of tables in the \from clauses, whereas Q13 required redistribution of {\groupby} attributes. All queries were refined with less than 4000 tokens. Given the current GPT-4o pricing of $\$2.5/million$ input tokens~\cite{chatgpt_citation}, even these ``hard-nut'' cases cost less than a cent apiece.

\begin{table}[!h]
\centering
{\small
\begin{tabular}{|p{3.8cm}|p{2.8cm}|p{0.8cm}|}
\hline
{\em Prompt Sequence} & {\em QID}& {\em \#Tokens}\\\hline\hline
IP, CCP, RCP & 1, 4, 5, 6, 9, 10, 16, 17, 21, 23, 24&\multirow{4}{*}{$<$ 4k}\\\cline{1-2}
IP, CCP, RCP, RCP& 3, 7, 8, 11, 12, 14, 15, 18, 22&\\\cline{1-2}
IP, CCP, RCP, CCP, CCP& 20 &\\\cline{1-2}
IP, CCP, RCP, CCP, RCP, RCP, {\bf CS}& 2, 13 &\\\hline
\end{tabular}
}
\caption{Prompt Sequences for Query Synthesis (E-TPCH)}
    \label{tab:prompt_count}
    \end{table}

\subsubsection{Effectiveness of Guidelines}
We observed that the LLM often deviates from the synthesis guidelines, especially with regard to complying with \QS on its provable extractions. Hence, feedback prompts were generated based on the output query to forcefully instantiate guidelines. For instance, the \QS for Q17 has tables {\sc Part, Web\_Lineitem w1, Web\_Lineitem w2}. However, the LLM-refinement uses two instances of {\sc Part}, and requires an explicit prompt {(\em Strictly use the tables as per \QS:} {\sc Part} table once, {\sc Web\_Lineitem} table twice) to prevent it from continuing to do so. The opposite phenomenon was seen for Q24, where multi-instance tables in \QS appeared only once in the synthesized \from clause. However, despite such issues, repeated feedback prompting was sufficient to eventually resolve all 23 queries.

\noindent\begin{figure}[!h]
    \centering
\begin{tabular}{cc}
  \includegraphics[width=0.5\linewidth]{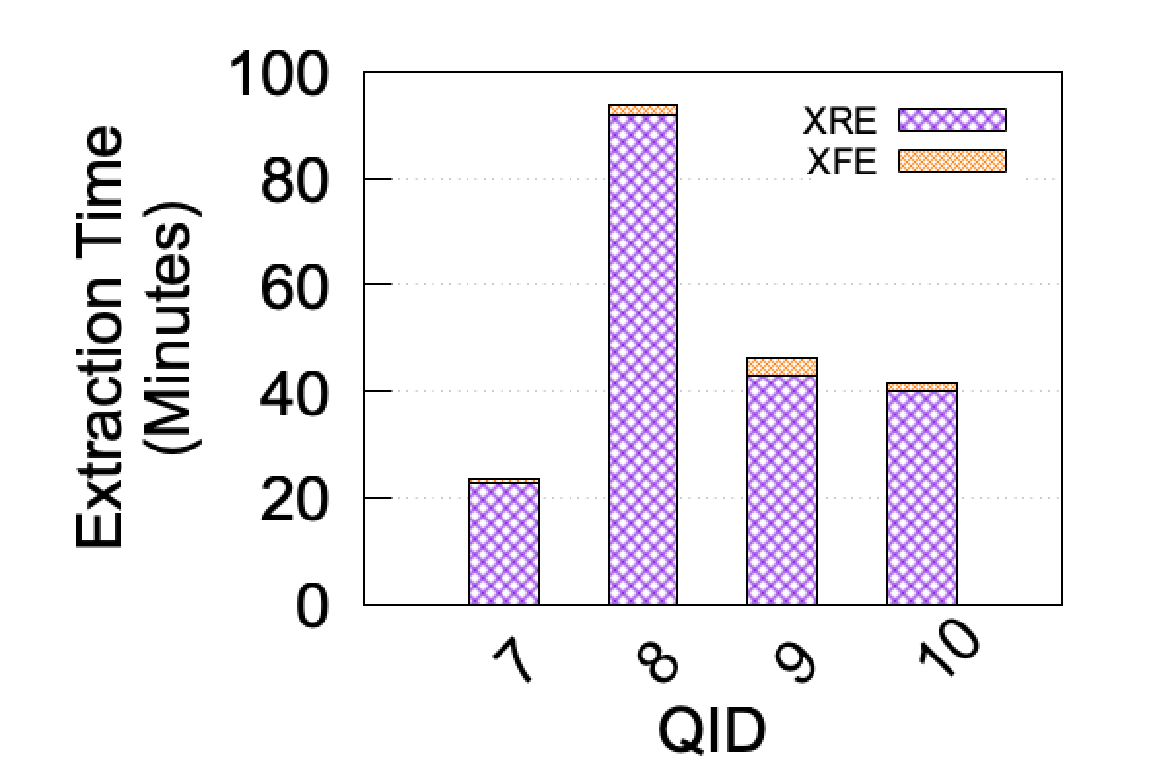}   & \includegraphics[width=0.4\linewidth]{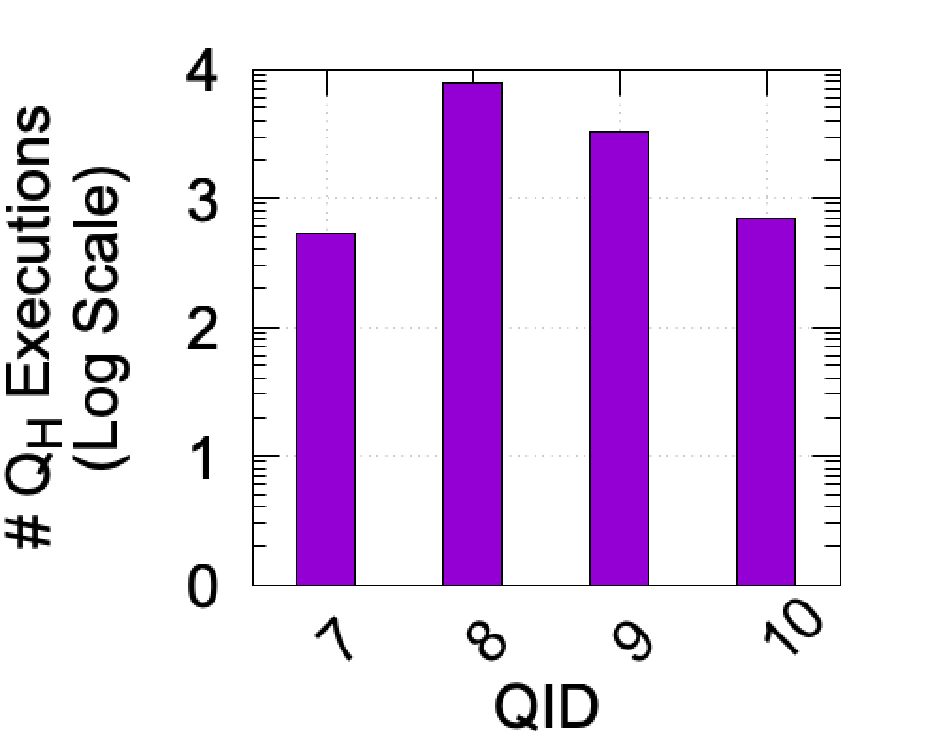} \\
  \multicolumn{2}{c}{(a) Extraction Times \hspace*{1.5cm}  (b) \QH invocations}\\
\end{tabular}
    \caption{Extraction Overheads (STACK)}
    \label{fig:stack}
\end{figure}

\subsection{STACK Extraction}
Our final experiment is on the 4 bi-directional STACK queries. All these extractions could be verified using \xdata.

The overheads incurred in the extractions are shown in Figure~\ref{fig:stack}. We observe that the extraction times are now much larger, with Q8 going up to almost two hours. The reason for this extended time is that the STACK database is large ($\sim$100 GB) and therefore minimization itself, rather than extraction per se, becomes an expensive exercise. Moreover, multiple rounds of such minimization are required to extract all the disjunction predicates (there are 5 constants in Q8's \sqlin predicate).

We also observe that in this benchmark, virtually all the time is taken XRE relative to XFE. This is because XRE handles the self-joins and disjunctions which take the primary computational effort. Whereas XFE's role is confined to nested structures connected by existential operators, which were found within a couple of feedback prompting iterations.

%% file: disc.tex
\section{The Road Ahead}
\label{sect:discuss}
At this stage, we have shown that it is indeed feasible, with a judicious combination of reverse and forward engineering, to extract substantively complex hidden SQL queries. But there remains ample opportunities for taking these ideas further, as discussed below.

\subsubsection*{Multi-level Nesting}
Our current extraction scope primarily covers queries with two levels of nesting. This is mainly due to GPT-4o's bias toward synthesizing unnested queries. Moreover, 
business logic requiring more nesting levels are often abstract, as in TPC-DS, which hampers XFE in deriving such structures. We are exploring whether it would be feasible to extend XRE from its current flat avatar to handle at least one level of nesting, thereby enhancing \utwo's overall scope for multi-level query hierarchies.

\subsubsection*{Multi-block Queries}
Our focus here was on {\tpch} and its derivatives, but queries in the TPC-DS benchmark~\cite{tpcds} pose novel challenges to {\utwo}. This is because they have multi-block structures constructed via multiple CTEs, instantiated many times, which makes identifying their signatures difficult as compared to base tables. Further, we found their high-level descriptions, as per the official documentation, to be insufficient for XFE to recognize the block boundaries. In our future work, we intend to study these issues with new strategies -- for instance, chain-of-thought prompting~\cite{cot} may help to address identification of block boundaries.

\subsubsection*{Multi-query Applications}
Thus far, we had only discussed applications with monolithic SQL queries embedded within them. But enterprise applications may comprise several queries. {\utwo} can be used in such multi-query scenarios if the individual queries are wrapped within separate functions~\cite{unmasque_sigmod}. In practice, modular application codebases typically have such wrappers. Moreover, using utilities such as \emph{GDB}~\cite{gnu_gdb} or \emph{objdump}~\cite{gnu_objdump}, individual functions can be isolated from the application binary~\cite{stackoverflow_extract_functions_elf_2025}. They can then be independently executed from their machine codes~\cite{stackoverflow_executing_machine_code_2025, github_frodox_execute_machine_code_2025}. Therefore, we expect that \utwo could be made viable for such real-world black-box applications.

%% file: related_work.tex
\section{Related Work}
\label{sect:relwork}
Classical query reverse engineering (QRE) ~\cite{talos}, \cite{scyth}, ~\cite{fastqre}, ~\cite{regal_plus, regal},  ~\cite{squares}, ~\cite{patsql}, \cite{qbe_application},  ~\cite{cubes} uses only query synthesis, and hints on the ground truth ~\cite{cubes}. They are instances of the program-by-example paradigm~\cite{DBLP:conf/aplas/GulwaniJ17} hosted in the SQL-world. However, due to the large enumerated search space, the scalability of such systems is weak~\cite{unmasque_sigmod}. Moreover, synthesizing a query that matches the user intention (i.e. business logic) is not achieved in most cases~\cite{cubes}, and requires significant human interaction.
Lastly, synthesizing complex OLAP queries such as the TPCH benchmark, has not been previously achieved in the literature.

Query forward engineering is gaining pace in automation using the contemporary technology of LLM-based \texttosql tools~\cite{text_to_sql_paperswithcode}. However, the reach of such tools is limited to simple cases~\cite{nl2sqlnot,stockinger,DBLP:conf/edbt/MitsopoulouK25}. Moreover, they require unambiguous text, and expect restricted schema structures~\cite{stockinger}. The \texttosql benchmarks such as Bird~\cite{birdbench} and Spider~\cite{spider} comprise queries that correspond to user questions posed to get immediate answers from the database, i.e. they lack complexity. In fact, there is recent evidence that applying such tools to engineer OLAP queries is unlikely to be successful~\cite{Ma2025}, and it is concluded that a human-in-the-loop workflow is required when AI is used as the synthesis agent.  Our own experience also bears this out -- only 5 {\tpch} benchmark queries were correctly formulated by GPT-4o \cite{chatgpt_citation} directly from their respective business descriptions, despite prior training on the same benchmark.

%% file: conclusion.tex
\section{Conclusions}
\label{sect:conclusion}

Hidden Query Extraction is a QRE variant with several industrial use-cases. 
We presented {\utwo}, a non-invasive system for extracting hidden queries of the complexity seen in popular OLAP benchmarks. {\utwo} features a unique bi-directional engineering architecture, wherein forward engineering establishes the query skeleton while reverse engineering fleshes out the details. Moreover, our evaluation showed that many queries could only be extracted by harnessing together their complementary abilities. We also found that HQE for industrial-strength queries is a complex learning task requiring powerful LLM models and reasoning power.

The XRE module deterministically extracts the core operators of the opaque application to generate a rich seed query featuring unions, algebraic predicates and disjunctions. Subsequently, XFE uses an LLM and combinatorial query synthesis to refine the seed to match the hidden query. The extraction overheads were found to be reasonable for modestly-sized database instances, and in our future work, we plan to look into scalability improvements.